\magnification=1200
\hyphenpenalty=2000
\tolerance=10000
\hsize 14.5truecm
\hoffset 1.truecm
\openup 5pt
\baselineskip=24truept
\font\titl=cmbx12
\def\dert#1#2{{{d#1}\over{d#2}}}
\def\der#1#2{{\partial#1\over\partial#2}}


\def\derss#1#2{{\partial^2#1\over\partial#2^2}}
\def\rg{r_g}

\def\kes{\kappa_{es}}

\def\ref{\par\noindent\hangindent 20pt}
\def\refig{\par\noindent\hangindent 15pt}
\def\mincir{\raise -2.truept\hbox{\rlap{\hbox{$\sim$}}\raise5.truept
\hbox{$<$}\ }}
\def\magcir{\raise -4.truept\hbox{\rlap{\hbox{$\sim$}}\raise5.truept
\hbox{$>$}\ }}
\def\rho{\varrho}
\def\mdot{\dot m}

\def\Jb{J_{\nu}}
\def\Hb{H_{\nu}}
\def\Kb{K_{\nu}}
\def\Lb{L_{\nu}}
\null\vskip 1.truecm
\centerline{\titl{GENERAL RELATIVISTIC RADIATIVE TRANSFER}}
\medskip
\centerline{\titl{IN HOT ASTROPHYSICAL PLASMAS:}}
\medskip
\centerline{\titl{A CHARACTERISTIC APPROACH}}
\vskip 1.5truecm
\centerline{Silvia Zane$^1$, Roberto Turolla$^2$, Luciano Nobili$^2$ 
and Myris Erna$^2$}
\medskip 
\centerline{$^1$International School for Advanced Studies, Trieste}
\centerline{Via Beirut 2--4, 34014 Miramare--Trieste, Italy}
\medskip 
\centerline{$^2$Department of Physics, University of Padova}
\centerline{Via Marzolo 8, 35131 Padova, Italy}
\vfill\eject

\beginsection ABSTRACT

In this paper we present a characteristic method for solving the 
transfer equation in differentially moving media in a curved spacetime. 
The method is completely general, but its capabilities are exploited at best
in presence of symmetries, when the existence of conserved quantities 
allows to derive analytical expressions for the photon trajectories in 
phase space. In spherically--symmetric, stationary configurations the solution
of the transfer problem is reduced to the integration of a single ordinary
differential equation along the bi--parametric family of characteristic rays.
Accurate expressions for the radiative processes relevant to continuum transfer
in a hot astrophysical plasma have been used in evaluating the source term,
including relativistic e--p, e--e bremsstrahlung and Compton 
scattering. A numerical code for the solution of the transfer problem in 
moving media in a Schwarzschild spacetime has been developed and tested. Some
applications, concerning ``hot'' and ``cold'' accretion onto non--rotating
black holes as well as static atmospheres around neutron stars, are presented
and discussed.

\bigskip\noindent
Subject headings: {\it accretion, accretion disks\/} \ -- \ {\it numerical 
methods\/} \ -- \ {\it radiative transfer\/} \ -- \ {\it relativity\/} 
 
\vfill\eject

\beginsection 1. INTRODUCTION

Radiative transfer in high energy, fast moving plasmas in a strong 
gravitational field is today at the basis of a large number of currently
interesting astrophysical applications; accretion onto compact objects, jets,
stellar collapse and supernova expanding envelopes are just some examples
of this. 

Since the pioneering works by Thomas (1930), Simon (1963) and Lindquist (1966),
astrophysical relativistic transfer received wide attention (see e.g. Mihalas
\& Mihalas 1984 for references to earlier papers). It was realized long ago
(see Castor 1972, Mihalas 1980 and references therein) that, for relativistic
flows, the interaction between matter and radiation is most easily described
if the material properties and the radiation field are evaluated in the 
frame in which the medium is at rest. The comoving frame transfer equation 
(CTE) 
has been considered by Mihalas (1980), Hauschildt \& Wehrse (1991) in the
framework of special relativity, and by Schmidt--Burgk (1978), Thorne (1981), 
Schinder \& Bludman (1989) in the general--relativistic case.
Different approaches for the solution of the relativistic transfer problem
in planar or spherical geometry
have been suggested. They can be grouped, schematically, into three
wide classes: direct solution of the CTE using discretization techniques, 
moment expansion and integration of the CTE along characteristic directions.

The solution of the CTE by finite differencing, like in the DOME method 
(Hauschildt \& Wehrse 1991), works well in geometrically thin layers, but
the treatment of extended atmospheres requires a prohibitive number 
of discrete elements to obtain a fair angular resolution. In the 
relatively simple case examined by Hauschildt \& Wehrse, the numerical
calculations must be performed on supercomputers even for low resolution grids.

The expansion of the specific intensity 
in spherical harmonics (moments) has the main advantage of reducing the 
dimensionality of the problem since the angular dependence is suppressed.
On the other hand the solution of the transfer problem is reconduced
to the solution of a recursive system of partial differential equations that
must be truncated at a given order, introducing a certain number of closure
conditions. This approach is at the basis of the flux limited diffusion theory
(FDT) developed by Levermore \& Pomraining (1981) and generalized by
Pomraning (1983), Anile \& Sammartino (1989) and Anile \& Romano (1992).
Although in the gray case the FDT provides a self--consistent closure function
by solving the differential equation for the flux limiter, the extension to the
frequency--dependent problem seems far from being obvious.
A very sophisticated, general--relativistic version of the moment formalism was 
presented by Thorne (1981). It is based on an expansion in projected, 
symmetric, trace--free (PSTF) moments and, upon truncation, the resulting 
system of equations can be solved introducing the required number of closure
conditions. The closure functions must be specified ``a priori'' and should 
reproduce the correct asymptotic limits for the radiation moments when 
free streaming and diffusion are approached. This method has been 
fruitfully  applied to the solution
of astrophysical problems with planar or spherical symmetry, both in the 
gray and
in the frequency--dependent case (see e.g. Turolla \& Nobili 1988, Nobili,
Turolla \& Zampieri 1993, Zampieri, Turolla \& Treves 1993). While the
arbitrariness of the closure functions is not a serious problem in the
gray case, where a large number of moments can be used, it becomes a major
complication when frequency--dependent transfer is tackled. In
fact, to make the numerical solution affordable, only the first two moments
can be taken into account, so that the choice of the closures has a 
non negligible impact on the results. Moreover, the extension to the 
bidimensional case is very complicated and requires the specification of
12 closure relations, making the method unacceptably dependent on the choice 
of a large number of free functions. 

Characteristics methods are based on the fact that the transfer equation is
just the Boltzmann equation for the photon distribution function in 
phase--space (see e.g. Lindquist 1966). The hyperbolic character of the 
Boltzmann equation implies that the CTE can be always reduced to a single
ordinary differential equation along the characteristic rays. The tangent ray 
method (TRM) developed in a series of papers by Mihalas and coworkers 
(Mihalas, Kunasz \& Hummer 1975, 1976a, b, Mihalas 1980) uses a 
semi--characteristic approach in which the integration is performed along
the characteristics of the ``spatial'' part of the differential operator
(the tangent rays), while the frequency derivative is treated by means of 
a standard finite--differences scheme. 
A fully characteristic method for the solution of the general--relativistic 
transfer problem has been discussed by Schmidt--Burgk (1978), 
Schinder (1988) and Schinder \& Bludman (1989). All these investigations 
dealt with stationary, spherically symmetric space--times, which admit
three Killing vectors: the existence of the associated constants of the motion
can be used to obtain simple expressions for the characteristic rays.
The analysis by 
Schinder \& Bludman (1989) was 
actually restricted to a spacetime characterized by
a stationary lagrangian line--element, which corresponds to a vanishing 
eulerian velocity field for the matter configuration; their test models
refer, in fact, to a static atmosphere. The work by 
Schmidt--Burgk (1978), although finalized to accretion onto a 
Schwarzschild hole, is, to our knowledge, the only example of an exact 
solution of the CTE taking into account both the effects of dynamics and 
strong gravity. 
For their simple mathematical structure, characteristic methods seem to be 
promising to cope with realistic astrophysical problems. Moreover, 
they can be quite naturally extended to more than one spatial 
dimension, the major complication coming from the higher number of ODEs that 
must be solved to compute the characteristic trajectories. 

Previous 
investigations were mainly concerned with the development of efficient 
methods for the solution of the CTE, 
assuming rather simple, often ``ad hoc'', expressions for 
the emission and absorption 
coefficients. This approach is completely justified if one is interested 
in investigating the formation of particular spectral features, like lines or
absorption edges. 
On the other hand, in all situations in which attention is focussed on the 
continuum, an accurate 
treatment of all relevant radiative processes becomes important. 
When dealing with hot plasmas,  
the dominant radiative processes are non--conservative scattering and 
bremsstrahlung. 
Solutions presented by Schmidt--Burgk (1978) refers to a hot, magnetized 
plasma and takes into account scattering absorption and synchrotron 
absorption/emission; the collisional term in the Boltzmann equation 
is written using suitable approximations.
On the other hand, 
approaches based on moment expansion (see e.g. Pomraining 1973 and 
references therein, Thorne 1981, Prasad {\it et al.\/} 1988)
do not permit an exact description of anisotropic and non--coherent 
scattering, usually treated in the Fokker--Planck 
approximation. A rigorous treatments of the Compton 
scattering can be found in Kershaw, 
Prasad \& Beason (1986), Kershaw (1987), 
Shestakov, Kershaw \& Prasad (1988), but their 
results are never been included in transfer codes devoted to  
astrophysical applications.

In the following we discuss a fully characteristic approach to 
the solution 
of the transfer equation in its more general form; results 
are then 
specialized to stationary, spherically--symmetric or 
plane--parallel configurations. 
Particular care will be devoted to a detailed treatment of the source 
term for an unmagnetized, fully ionized, 
non--degenerate hydrogen gas. A numerical code is described and 
applications to accretion onto black holes 
and neutron stars are finally presented. 

\beginsection 2. RADIATIVE TRANSFER 

In this section we consider the characteristic form of the 
radiative transfer equation in the comoving frame. 
In subsection a) the CTE and the equations for its characteristic 
trajectories are derived in the more general case, when no symmetries 
are present. A particularization to spherically--symmetric 
configurations is presented in subsection b). Finally in 
subsection c) the choice of boundary conditions is discussed. In 
relativistic transfer the radiation field is naturally described 
by the photon distribution function in the phase--space, $f$, that is related 
to the specific intensity by $ 2f = c^2 I/h^4 \nu^3$.  
Geometrized units ($c \, = \, G \, =\, h\, = \, 1$) are used throughout and 
lengths are in units of the gravitational radius $r = 2M$. 

\medskip
\centerline{\it a) The Radiative Transfer Equation}
\medskip
The relativistic transfer equation, written in covariant form, is just the
Boltzmann equation  for $f({\bf x, p})$
$$\dert{f}{\xi} = g({\bf x,p})\eqno(1)$$
where ${\bf p}\equiv d{\bf x}/d\xi$ is the photon 4--momentum, $\xi$ is
an affine parameter along the null geodesic and the collisional term $g$ 
accounts for the 
interactions between matter and radiation (see e.g. Lindquist 1966, 
Thorne 1981). The 
differential operator in equation (1) acts not merely in spacetime but
in the full photons phase--space, made up by the spacetime plus the null
tangent space at each point along the photon trajectory. 

Since $f({\bf x,p})$ is a relativistic invariant, equation (1) holds in
any frame. However, the material properties (e.g. opacity and emission 
coefficients, scattering cross--section), which enter the expression of
the source term $g$, are naturally defined with respect to observers who
are locally and instantaneously at rest with the matter (LRF). 
In the following we adopt a fiducial observer comoving with the fluid, which
carries a tetrad ${\bf e}_{\hat a}$ and has 4--velocity ${\bf u}\equiv 
{\bf e}_{\hat 0}$. If spacetime, matter and radiation share some 
common symmetries, the orientation of some of the spatial vectors of the 
tetrad follows in a natural way. For example, in spherical symmetry, as it will
be discussed in detail later on, it is convenient to chose 
${\bf e}_{\hat 1}$ orthogonal to the $\theta$ and $\phi$ 
coordinate directions. With respect to the tetrad, the components of the photon 
4--momentum are

$$p^{\hat a} = (E,E\mu,E(1-\mu^2)^{1/2}\cos\Phi,E(1-\mu^2)^{1/2}\sin\Phi)\,
\eqno(2)$$
where $E$ is the photon energy, $\mu$ is the cosine of the angle between the
photon direction and ${\bf e}_{\hat 1}$, and $\Phi$ is the corresponding 
azimuthal angle, all measured in the LRF. The three quantities $E$, $\mu$
and $\Phi$ have an immediate physical meaning and they will be used as
independent variables (momentum variables) together with the spacetime 
coordinates $x^i$ to tick events on the light--cone of the
phase--space. The total derivative in equation (1) can be explicitated as
$$\eqalign{
& \der{f}{x^i}p^i + \der{f}{p^{\hat a}}\dert{p^{\hat a}}{\xi} =  \cr
& \der{f}{x^i}p^i + \der{f}{E}\dert{E}{\xi} + \der{f}{\mu}\dert{\mu}{\xi}
+ \der{f}{\Phi}\dert{\Phi}{\xi} = g\, , \cr}
\eqno(3)$$
where $p^i = p^{\hat a}e^i_{\hat a}$. The variation of $E$, $\mu$ and $\Phi$
along the photon trajectory can be obtained from the equation of the
null geodesic, written in terms of $p^{\hat a}$
$$\dert{p^{\hat a}}{\xi} + \Gamma^{\hat a}_{\hat b \hat c}p^{\hat b}p^{\hat c}
= 0\, ,\eqno(4)$$
where $\Gamma^{\hat a}_{\hat b \hat c} = e_i^{\hat a}e^j_{\hat b}e^i_{\hat c;
j}$ are the Ricci rotation coefficients. Recalling the expression of $p^{\hat
a}$ given in equation (2), we finally get

$$\eqalignno{
& \dert{E}{\xi} = -\Gamma^{\hat 0}_{\hat b \hat c}p^{\hat b}p^{\hat c}
& (5a)\cr
& \dert{\mu}{\xi} = -{1\over E}\left(
\Gamma^{\hat 1}_{\hat b \hat c} - \mu\Gamma^{\hat 0}_{\hat b \hat c}
\right)p^{\hat b}p^{\hat c} & (5b)\cr
& \dert{\Phi}{\xi} = -{1\over{E(1-\mu^2)^{1/2}}}\left(
\cos\Phi\Gamma^{\hat 3}_{\hat b \hat c} -\sin\Phi\Gamma^{\hat 2}_{\hat b 
\hat c}\right)p^{\hat b}p^{\hat c}\, . & (5c)\cr}$$
We note that all the information about the spacetime curvature and the
flow dynamics are contained in the tetrad field and enter the Boltzmann 
equation via the tetrad vectors themselves and their local rates of change 
which appear in the Ricci coefficients.

Equation (3), together with the set (5), is the more general form of the 
transfer equation and holds for arbitrary flow motions in any given 
spacetime. In the next subsection we will discuss how the existence of
spacetime symmetries implies that
the distribution function is independent on some of the phase--space variables,
easing the solution of the transfer problem.
\medskip
\centerline{\it b) Transfer in Spherically--symmetric Spacetimes}
\medskip

Let us consider the more general spherically--symmetric spacetime, described,
in spherical coordinates, by the line--element
$$ds^2 = g_{00}(r,t)dt^2 + g_{11}(r,t)dr^2 + r^2(d\theta^2 + 
\sin^2\theta d\phi^2)\, .$$
Spherical symmetry implies that there exist two constants of the motion,
$L_z$ and $L$, which are related to the components of the photon
4--momentum by $L_z = p_3 = r^2\sin^2\theta p^3$, $L^2 = r^4[(p^3)^2\sin^2
\theta + (p^2)^2]$. These two expressions take a very simple form,
and lead to a major simplification in the transfer equation, if
the fluid configuration and the radiation field are themselves 
spherically--symmetric. In this case the spatial 3--velocity ${\vec v}$ of the 
comoving 
observer, measured by the stationary observer $\delta_0^i/\sqrt{-g_{00}}$, 
is in the radial direction and the most convenient choice for the tetrad is
$$\eqalign{ 
            & e_{\hat 0}^i = \left( {\gamma\over{\sqrt{-g_{00}}}},
              {{\gamma v}\over{\sqrt{g_{11}}}},0,0\right) \cr
            & e_{\hat 1}^i = \left( {{\gamma v}\over{\sqrt{-g_{00}}}},
              {\gamma\over{\sqrt{g_{11}}}},0,0\right) \cr
            & e_{\hat 2}^i = (0,0,r^{-1},0)\cr
            & e_{\hat 3}^i  = (0,0,0,r^{-1}\sin^{-1}\theta)\cr}\eqno(6)$$
where $\gamma = (1-v^2)^{-1/2}$.
The constants $L_z$ and $L$ may be then expressed in terms of the tetrad 
components $p^{\hat a}$ as 
$$\eqalignno{
& L_z = L\sin\Phi\sin\theta & (7a)\cr
& L^2 = r^2E^2(1-\mu^2)\, . & (7b)\cr}$$
In spherical symmetry,
the photon distribution function must be independent on both the polar angles
$\phi$ and $\theta$. Since, from equation (7a), we have $\Phi = 
\Phi(\theta)$, it follows that isotropy in coordinate space implies also that 
$\partial f/\partial\Phi = 0$ and the Boltzmann equation reduces to
$$\der{f}{t}p^0 +\der{f}{r}p^1 + \der{f}{E}\dert{E}{\xi} + \der{f}{\mu}
\dert{\mu}{\xi} = g\, . \eqno(8)$$

In the further hypothesis that the spacetime is stationary, the existence
of a time--like Killing vector provides a third conserved quantity, $p_0\equiv
-E_\infty$, which can be used to obtain a simple expression for the photon 
energy along each ray in the LRF
$$E = {{E_\infty}\over{y(1 + \mu v)}}\, ;\eqno(9)$$
in the previous expression $y=\gamma\sqrt{-g_{00}}$ is the
specific energy of the fluid, as measured by a static observer at infinity.
Clearly the differential operator in the 
transfer equation is Pfaffian, so it is always possible to solve the Boltzmann 
equation along its characteristic directions, i.e. 
along the photon trajectories in the 7--dimensional phase--space. 
In the case at hand, these
trajectories lie in a 4--dimensional hypersurface and can be 
obtained solving equations (5a) and (5b) together with 
$$
\eqalignno{
\dert{t}{\xi} & = p^0 = -{{E_\infty}\over{g_{00}}} & (10a) \cr
\dert{r}{\xi} & = p^1 = {{E y}\over\rg} (\mu + v ) \, . & (10b)\cr}
$$
Actually, the existence of the two constants of motion $L$ and $E_\infty$ 
yields analytical expressions for both $\mu$ and $E$, as functions of $r$,
along each photon trajectory:
$$
\eqalignno{
\mu & = {-y^2 v b^2 \pm r \left ( r^2 + b^2g_{00} \right)^{1/2} \over 
r^2 + b^2 y^2 v^2  } & (11a) \cr  
E & = { {b^2 y^2 v^2 + r^2}  \over { \left [r^2 \pm r v \left ( r^2 + b^2 g_{00}
\right )^{1/2} \right ] y }}E_{\infty} & (11b) \cr} 
$$
where the impact parameter $b = L/E_\infty$ has been introduced. 
Due to spherical symmetry, only positive $b$'s need to be considered,
negative values of the angular momentum give exactly the specular picture,
so in the following $b^2$ will be used as a parameter. 
It can be easily shown that the plus/minus sign in equations (11) 
refers to photons for 
which $\mu + v$ is always positive/negative. This implies, see equation (10b),
that the radial coordinate is always increasing/decreasing along the
path and that the condition $\mu + v =0$ defines the locus of turning points
for the trajectories. This is just a manifestation of aberration: the turning 
points, in fact, are located where the cosine of the angle between the photon
and radial directions, measured by the stationary observer, vanishes. 

Specializing to the vacuum Schwarzschild solution, photon 
trajectories in physical space may be divided into three classes (see e.g.
Misner, Thorne \& Wheeler 1973): a) 
those connecting radial infinity with the event horizon, characterized by
impact parameters in the range $0\leq b^2< 27/4$; b) those that are trapped
in the region $1\leq r<3/2$ and c) those for which it is always $r>3/2$. 
Trajectories of the latter two types have $b^2>27/4$. 
The limiting value $b^2 = 27/4$ corresponds to the circular photon orbit.
The plot of $\mu = 
\mu(r;b)$ and of $E/E_\infty = \epsilon (r;b)$ is shown in figures 1a and 1b
for a free--fall velocity law, $u^1 = r^{-1/2}$. 
As can be seen from the the figures, photons starting at the horizon 
can reach infinity with non--zero energy 
only if they are emitted exactly in the radial 
direction ($\mu = 1$) with an infinite energy, while 
ingoing photons that leave infinity 
with zero angular momentum reach the horizon halving their initial energy. 
At large values of $r$ 
all rays concentrate at $\mu = \pm 1$, as radial streaming is approached.  
Trajectories with an impact parameter 
equal to the critical value $b^2 = 27/4$ exhibit a 
saddle point at $r = 3/2$. 

In the following we will concentrate on the case in which both matter and the 
radiation field are stationary.
Under this assumption the 
distribution function depends only on three variables, $r$, $E$ and $\mu$, and
since it is $E= E(r)$, $\mu = \mu(r)$ (see equations [11a], [11b]), the
radial coordinate itself can serve as a (non--affine) parameter along the 
null geodesics.
The Boltzmann equation can be then integrated in the domain of existence of 
each photon trajectory. This particular choice appears to be 
convenient for a number of reasons, although it poses some 
numerical problems, as it will be discussed later on. 
First of all, the treatment of boundary conditions is much simpler when 
the radial coordinate is the independent variable and this 
avoids also the integration of equation (10b) along with the transfer equation. 
Moreover, when scattering is taken into account, the source term depends on 
the integrals of $f$ over angles, which must be evaluated at both 
constant $r$ and $E$. The knowledge of $f(r)$ avoids the use of 
spline or other interpolation algorithms, which is time--consuming and would
be needed in the case of a different parametrization of the photon 
trajectories. In conclusion, 
at least for what concerns the radiation field, the transfer problem 
can be solved integrating numerically the single differential 
equation
$$
\dert{f}{r} = {\rg \over {y (\mu + v )}}{g \over E} \eqno(12)
$$
for different values of the two parameters $b$ and $E_\infty$.

At variance with what happens using other methods, like for example 
expansion in PSTF moments, 
this kind of approach makes a great simplification in 
the mathematical structure of the 
problem: 
in fact, the non--grey problem can be solved without 
integration of complicated systems 
of partial differential equations. 
Moreover, no closure is needed and 
this formalism gives as result the full radial, frequency and angle 
dependent solution. As we will discuss in detail in the next section, is just 
the knowledge of the angular dependence of the distribution function, 
lost when the moments of $f$ are used as dependent variables, that gives 
the possibility to use the characteristic rays method to study 
the Compton scattering in its more general form; this approach 
naturally preserves the hyperbolic character of the Boltzmann equation.  

In this investigation we focus our attention on the calculation of the
radiation field and, thus, we restrict
our discussion to the case in which velocity, density 
and temperature profiles are fixed {\it a priori\/}, similarly to what was
done by Mihalas (1980) in the special relativistic
case and Schinder \& Bludman (1989) in the general relativistic, static case.
Clearly, the full solution of the radiation hydrodynamical problem 
requires the simultaneous integration of the transfer equation together with
the Euler, continuity and energy equations that, in turn, depend on the gray 
mean intensity $J$ and on the gray radiative flux $H$:
$$
J = {1 \over 2} \int_0^{\infty} d \nu
\int_{-1}^{1} I  d \mu 
= {1 \over 2} \int_{0}^{\infty} d \nu \int_{-1}^{1} f  \nu^3 d \mu
$$
$$
H = {1 \over 2} \int_0^{\infty} d \nu
\int_{-1}^{1} I \mu  d \mu 
= {1 \over 2} \int_{0}^{\infty} d \nu \int_{-1}^{1} f  \nu^3 \mu d \mu\, .
$$
The coupled solution of the transfer and gasdynamical equation poses, 
therefore, the same difficulty encountered in the integration of the transfer 
equation alone in presence of scattering. A numerical technique for the
solution of the integro--differential scattering equation is discussed in
section 4a. The same method can be applied to the full radiation 
hydrodynamical problem and an example is presented in section 5c.

\medskip
\centerline{\it c) Boundary Conditions}
\medskip
Because there is not a one--to--one map between $r$ and $\xi$,
equation (12) must be integrated twice for each value of $b^2$,
in correspondence with the two solutions for $\mu$ and $E$ given by
equations (11a), (11b). At the same time, two different boundary conditions 
for the distribution function $f$ must be imposed, taking into account that
the plus (minus) sign in equations (11) corresponds to outgoing (ingoing)
trajectories. The boundary condition for ingoing characteristics of type a) 
is prescribed 
in the standard way: for a non--illuminated atmosphere, for example, it is
just $f=0$ at the outer edge of the integration domain. This is also the only 
condition required to integrate the transfer equation along all 
characteristics of type c), since integration can be started at large $r$ with, 
say, $f=0$ and carried out until the turning point is reached storing the 
computed value of $f$, which is then used as the initial condition along the
outgoing branch of the trajectory. The remaining rays, including 
characteristics
of type b), can be treated much in the same way if there exists a region in
the flow where the effective depth $\tau_{eff}$ becomes larger than unity at 
any frequency and LTE is
attained. In this case, in fact, the required boundary condition is simply
$f = B_\nu (T)/E^3$, $B_\nu (T)$ is the Planck function at temperature $T$,
at a radius $\overline r$ such that $\tau_{eff}(\overline r)>1$. 

Although this
is the standard case for stellar atmospheres, including accretion flows onto
compact stars, a different situation may arise when dealing with accretion
onto black holes: for low values of the accretion rate, for example, the 
flow is 
optically thin all the way down to the horizon (see e.g. Nobili, Turolla \& 
Zampieri 1991). Now a boundary condition for $f$ must be imposed at $r=1$
for rays starting at the event horizon. Since $E$ goes to infinity there,
both the distribution function and $g$ must vanish. The product $E(\mu + v)$, 
however, does not vanish for all outgoing rays at the horizon, so $g=0$ implies
also $df/dr = 0$. In order to avoid numerical overflows, integration is
started at a radius $r_{in}$ fractionally larger than unity, with the 
regularity condition $df/dr = 0$.
The two rays with $b^2 = 27/4$ are peculiar since they intersect at $r=3/2$ 
(the saddle point) which is also a critical 
point for equation (12). We still integrate the transfer equation along 
these particular rays taking as a regularity condition $g =0$ at $r = 3/2$. 
Strictly speaking, this condition is exact only in the case in which 
the effective optical depth is larger than unity at the last photon orbit; in 
other cases there is no physical reason to ask for thermalization and 
the value of $f$ may be undetermined. However, 
since the radial derivative of $f$ diverges at the critical point, we found 
that, in a finite differences numerical scheme, the solution of 
the differential equation fast relaxes and the final result 
is probably not strongly affected by the value of the distribution 
function at $r=3/2$.


\beginsection 3. THE SOURCE FUNCTION

In the following we deal with an unmagnetized, fully ionized, 
non--degenerate hydrogen gas in which emitters and absorbers are in local 
thermal equilibrium at a temperature $T$. We consider also 
the case in which electrons are relativistic ($T \magcir 5\times 10^9$ K), 
and present a fully general treatment of Compton scattering.
However, for the sake of simplicity, we focus our attention only on 
thermal emission and absorption together with scattering from free electrons; 
other processes, as pair production and double Compton scattering, that may
be relevant at such high temperatures, are outside the scope of this paper.
In this section physical units are used; 
$\gamma$ and $\tau$ denote the dimensionless 
photon energy and electron temperature, both in units of $m_ec^2$; $K_p
(x)$ is the modified Bessel function of the second kind. 
\medskip
\centerline{\it a) Thermal bremsstrahlung}
\medskip
The source term for spontaneous emission and absorption, including
stimulated emission, can be written as
$$
g = {\eta \over 4 \pi hc E^2 } - \chi E f\, ,
\eqno(13)
$$
where $\eta $ and $\chi$ are the emission and absorption coefficients, 
measured in the comoving frame. Because of the assumed equilibrium,
Kirchhoff law yields:
$$
{\eta \over 4 \pi \chi } = B_{\nu} \left ( T \right )
$$
with $\nu=E/h$.
In the medium we are considering, the dominant true
emission and absorption processes are electron--proton and electron--electron
bremsstrahlung; in the following we will indicate as $\kappa_{ff}$ the 
correspondent total opacity. 
The free--free contribution to the source term is then
$$
{g_{ff} \over E} = \rho \kappa_{ff}\left ({ B_{\nu} 
\over hcE^3 } - f  \right ). \eqno(14)
$$ 

The photon spectrum from bremsstrahlung is usually described in terms of the 
velocity--averaged Gaunt 
factor $G$; in the non relativistic regime tables for $G$ have been presented 
by Karzas \& Latter (1961). 
However, as discussed by Gould (1980), contributions to the 
total energy loss rate due both to relativistic 
corrections in the electron velocity distribution and to  
electron--electron bremsstrahlung are already of order 10 \% 
at $T\sim 10^8$ K become as large as 30 \% at $T\sim 10^9$ K.
Free--free emissivity from a relativistic thermal plasma has been investigated 
by several authors (see e.g. Alexanian 1968; Quigg 1968; Haug 1975; Gould
1980; Stepney \& Guilbert 1983; Dermer 1984, Dermer 1986). 
The photon spectrum from e--p emission involves a single 
quadrature over the relative Lorentz factor of the interacting particles 
$\gamma_r$ (see
e.g. Dermer 1986)
$$\eta_{e-p}(\gamma,\tau) = {{n_en_pc}\over{\tau K_2(1/\tau)}}\int^\infty_{1
+\gamma}d\gamma_r\left(\gamma_r^2 -1\right)\dert{\sigma_{B-H}(\gamma,\gamma_r)}
{\gamma}\exp\left(-{{\gamma_r}\over{\tau}}\right)\, ,$$
where $d\sigma_{B-H}(\gamma,\gamma_r)/d\gamma$ is the Bethe--Heitler cross
section corrected for the Elwert factor (see e.g. Heitler 1936) and $n_e$,
$n_p$ are the number density of electrons and protons. The previous expression
holds for $\tau\ll m_p/m_e$, so that protons can assumed to be at rest in
the lab--frame. 

Electron--electron emissivity is more complicated since now both particles 
have the same mass and a quadrupole contribution appears. The standard 
expression involves a five--fold integral of the totally differential 
cross--section (Haug 1975), but, as shown by Dermer (1984, 1986), it can be
reduced to a triple integral exploiting the covariance of Haug's formula
to evaluate the cross--section in the CM--frame. The final result is
$$\eqalign{\eta_{e-e}(\gamma,\tau) = &
{{n_e^2c}\over{4\tau K_2^2(1/\tau)}}\int^\infty_1
d\gamma_r{{\left(\gamma_r^2 -1\right)}\over{\left[2\left(\gamma_r+1\right)
\right]^{1/2}}}\int^{\zeta(\gamma_r)}_0{{d\gamma^*}\over{\gamma^*}}
\dert{\sigma^*_{e-e}(\gamma^*,\gamma_r)}{\gamma^*}\times\cr
& \exp\left\{-
{{\left[2\left(\gamma_r+1\right)\right]^{1/2}}\over{\tau}}\left({{\gamma^2
+\gamma^{*2}}\over{2\gamma\gamma^*}}\right)\right\}\cr}$$
(see Dermer 1986 for notation). 

The numerical evaluation of both $\eta_{e-p}$ and $\eta_{e-e}$ poses no 
particular problems and has been carried out following Dermer (1986) in
the ranges $2\times 10^{-2}\leq\tau\leq10$, \  $2\times 10^{-2}\leq h\nu/KT\leq
25.12$. Numerical results for
the total Gaunt factor  
were then fitted with the analytical function (see Stepney
\& Guilbert 1983)
$$G=\cases{(A+Bx)\ln(1/x)+ C+ Dx,& $x=h\nu/KT\leq 2.51$\cr
\,& \ \cr
\alpha x^2+ \beta x+\gamma +\delta/x,& $x> 2.51$,\cr}$$
deriving, for each $\tau$, the set of coefficients $A,\ldots,\delta$. The
Gaunt factor can be then obtained at any value of $\tau$ and $h\nu/KT$
by means of a suitable interpolation/extrapolation. At temperatures below
$\sim 10$ keV $(\tau\mincir 0.01)$, the asymptotic limits of Gould (1980)
are used for both e--p and e--e emissivity.
\medskip
\centerline{\it b) Electron scattering}
\medskip

The second important radiative process we consider is scattering from free 
electrons: we recall that one of the major complications 
encountered in solving the transfer equation comes from its non--local 
character. In fact, even limiting to the coherent and isotropic case, the 
source term is
$$
{g_{es} \over E} = \rho\kes\left (j_{\nu} 
 - f  \right ),\eqno(15)
$$  
where $\kappa_{es}$ is the Thomson opacity and 
$$
j_{\nu} = {1 \over 2} \int_{-1}^{1} f \left (r,\mu, E \right ) d\mu
$$
is the zero--th moment of the distribution function. 
Allowing for the more realistic case of Thomson scattering,
the correspondent cross section has a monopole plus a quadrupole 
angular dependence (see e.g. Chandrasekhar 1960) yielding
$$
{g_{es} \over E} = \rho \kes \left [ { 3 \over 8}
\left [ \left ( 3 j_{\nu} - k_{\nu} \right ) 
- \mu^2 \left ( j_{\nu} - 3 k_{\nu} \right ) \right ]
 - f \right ] \, ,   \eqno(16)
$$
where
$$
k_{\nu} = {1 \over 2} \int_{-1}^1 f \left (r,\mu, E \right ) \mu^2 d \mu
\, .
$$

The Thomson limit can be assumed to correctly describe electron scattering 
when the energy exchange in a single collision can be
safely ignored. On the other hand, in high temperature regions 
non--conservative effects and quantum corrections play a fundamental role in
shaping the emergent spectrum. The derivation of the general expression for the
Compton source term can be found e.g. in Pomraning (1973) and is briefly 
outlined below, mainly to introduce some basic ideas which will be used later 
on when the numerical scheme is discussed. With reference to a single 
scattering, $\vec n$ denotes the incident photon direction
and $\xi = {\vec n} \cdot {\vec n'}$, where primed quantities refer to the 
scattered photon.
For an incident photon energy $\gamma$ and an electron velocity 
$ \vec v_e$, 
the Klein--Nishina formula gives the probability of scattering into the energy
$\gamma'$ and the direction $\vec n'$
$$
\eqalign{
\sigma \left (\gamma \rightarrow \gamma', {\vec n} \rightarrow {\vec n'}, 
\vec v_e \right ) = & {r_0^2 \over 2 \gamma \nu \lambda} \left \{
1 + \left [ 1 - { \left ( 1 - \xi \right ) \over \lambda^2 D D' }
\right ]^2 + { \left ( 1 - \xi \right )^2 \gamma \gamma' \over \lambda^2 D D'
} \right \} \cr 
& \times \delta \left [ \xi - 1 + \lambda {D \over \gamma' } - \lambda 
{D' \over \gamma} \right ]  \, , \cr } \eqno(17) 
$$ 
where 
$$
D = 1 - {\vec n} \cdot {\vec v_e}/c, \qquad D' = 1 - {\vec n'} 
\cdot {\vec v_e}/c, \qquad \lambda = \left ( 1 - {v_e^2 \over c^2} 
\right )^{-1/2} \, ,
$$
$r_0$ is the classical electron radius and $\delta$ is the Dirac delta 
function. Integration over the relativistic maxwellian distribution
$$
f_e(v_e) = { \lambda^5 \exp \left ( - \lambda/\tau \right ) \over 4 \pi \tau 
c^3 K_2 \left ( 1/\tau \right )} \eqno(18) 
$$
gives the Compton Scattering Kernel (CSK)
$$
\eqalign{
\sigma \left ( \gamma \rightarrow \gamma', \xi , \tau \right ) & = { 3 \over 16
\pi \gamma \nu } \int d {\vec v_e} {f_e \left (v_e \right ) \over \lambda} 
\left \{ 1 + \left [ 1 - { 1 - \xi \over \lambda^2 D D' } \right ]^2 + 
\right.\cr
& \left.{ \left ( 1 - \xi \right )^2 \gamma \gamma' \over \lambda^2 D D' } 
\right \}
\delta \left ( \xi - 1 + \lambda {D \over \gamma'} - \lambda {D' \over 
\gamma} \right )  \, .  \cr} \eqno(19)
$$
Here the CSK is normalized with respect to $\kappa_{es} \rho$, 
which is reciprocal of the Thomson mean free path;
the inverse probability, related to the scattering emissivity, can be obtained
from the detailed balance condition
$$
\sigma \left ( \gamma \rightarrow \gamma', \xi , \tau \right ) \gamma^2 \exp{ 
\left (-
\gamma / \tau \right )} =
\sigma \left ( \gamma' \rightarrow \gamma, \xi , \tau \right ) 
\gamma'^2 \exp{ \left ( -
\gamma' / \tau \right )} \, . \eqno(20)
$$

Further integrations over all outgoing photon directions and energies provide 
the source term appearing in the Boltzmann equation
$$
\eqalign{ {g_{C}
 \over E} = & \kappa_{es}
 \rho \int_0^{\infty} d \gamma' \int_{4\pi}
d \Omega' \left ( {\gamma' \over \gamma} \right )^2
\sigma \left ( \gamma' \rightarrow \gamma, \xi , \tau \right ) 
f \left (r, {\vec n'}, \gamma' \right ) \left [ 1 + 
{ f \left (r, {\vec n}, \gamma \right ) \over 2} \right] \cr
& - \kappa_{es} \rho 
\int_0^{\infty} d \gamma' \int_{4\pi}
d \Omega' 
\sigma \left ( \gamma \rightarrow \gamma', \xi , \tau \right ) 
f \left (r, {\vec n}, \gamma \right ) \left [ 1 + 
{f \left (r, {\vec n'}, \gamma'  \right ) \over 2} \right] 
\, . \cr }\eqno(21) 
$$
Inserting equation (20) into equation (21), the latter can be written in the 
more compact form
$$
\eqalign{
{g_C \over E} = & \kappa_{es} \rho \int_0^{\infty} d \gamma' \left [ 
\exp \left ( - { \gamma - \gamma' \over 2 } \right ) \left ( {f \over 2 } 
+ 1 \right ) - { f \over 2} \right ] \int_{4 \pi} d \Omega' 
\sigma \left ( \gamma \rightarrow \gamma', \xi, \tau \right ) f' \cr
& - \kappa_{es} \rho \sigma _{00} f \, , \cr } \eqno(22)
$$ 
where $f' = f(r, {\vec n'}, \gamma')$ and
$$
\sigma_{00} = \int_0^{\infty} d \gamma' \int_{4 \pi} d \Omega' 
\sigma \left ( \gamma \rightarrow \gamma' , \xi, \tau \right ) \eqno (23)
$$ 
is the zero--th moment of the CSK (Shestakov, Kershaw \& Prasad 1988). 
In the previous expressions, non--linear terms account for stimulated 
scattering. 

The general task of computing the moments of the CSK was undertaken
by Shestakov {\it et al.\/} (1988). They have shown that by performing
the integration over $\gamma'$ first and exploiting the $\delta$--function,
the expression of the zero--th moment, which is originally a fivefold 
integral, can be reduced, after a lot of non--trivial algebra, to a single 
quadrature
$$
\sigma_{00} =  { 1/ \gamma  \over 2 K_2 \left ( 1/ \tau \right ) } 
\int_{0}^{\infty} d z \gamma z \sigma_{0} \left ( \gamma z \right ) 
\exp \left [ - { 1 \over 2 \tau } \left ( z + { 1 \over z} \right ) 
\right ] \eqno (24a)
$$
where:
$$
\eqalign{
y \sigma_{0} \left ( y \right ) = & { 3 \over 8 y } \left [ 
{ y^2 - 2 y - 2 \over y } \ln \left ( 2y + 1 \right ) + 
{ 2 y^3 + 18 y^2 + 16 y + 4 \over \left ( 2 y + 1 \right )^2 } 
\right ] \cr
= & y \left ( 1 - 2 y + { 26 \over 5} y^2 - {133 \over 10} y^3 +
{ 1144 \over 35 } y^4 - \cdots \right )\quad {\rm for } \  |y| < 
{1 \over 2 } \, . \cr} \eqno (24b)
$$

The full evaluation of the Compton source term involves
a number of very complicated six dimensional 
integrals of the distribution function weighted by the CSK 
for each value of $\gamma$, $\tau$, $\mu$. Because only discrete values of 
the distribution function will be available, all the six quadratures should 
be, in principle, evaluated numerically
at each grid--point and this would make the integration 
of the transfer equation prohibitively time--consuming. However, as discuss by 
Kershaw, Prasad \& Beason (1986), two of the three integrals in the CSK become 
analytical if a particular polar axis for projecting the electron velocity
is chosen.
Moreover, Kershaw (1987) presented an efficient method 
for calculating the single integral of the CSK over $\gamma'$ or $ \xi$ and the 
double integral over both these variables. 
A detailed discussion of our algorithm for the evaluation of the 
first addendum in the Compton 
source term, that is essentially a re--adaptation of Kershaw's method, 
is presented later on.

Although the treatment we have just described is the more general to handle
Comptonization and proved to be reasonably fast, it remains very 
time--consuming, so it is useful to have approximated expressions of $g_{C}$ 
that can be used in some regimes.
As it is well known, 
the complicated nature of the CSK has led many authors to model the Boltzmann 
equation by a diffusion equation in the frequency space. This approach, 
the Fokker--Planck approximation, was firstly used 
by Kompaneets (1956) in the limit 
of small $ \gamma $ and $\tau$. Relativistic corrections to the Kompaneets 
equation can be included modifying the diffusion coefficient, and a number of 
efforts were devoted to extend its original form (Fraser 1966, as quoted
in Pomraning 1973, Cooper 1971). More recently, Prasad {\it et al.\/} 
(1988) derived an exact analytical expression for the diffusion coefficient 
that holds for arbitrary values of $\gamma$ and 
$\tau$, in the assumption of a nearly isotropic radiation field. 
The main simplification introduced by the Fokker--Planck approximation 
is that the integral operator in the transfer equation is replaced by an 
infinite order differential operator that, for small values of $\gamma $ and 
$\tau$, truncates at a finite order. 
The method, originally developed for the non--relativistic transfer 
equation, is based on an expansion of the specific 
intensity in a Taylor series about $\nu' = \nu$.
At the first order in $\gamma$ and $\tau$, Fraser's result is
$$
\eqalign{
g_{C}E^2 = & - \kappa_{es} \rho \left ( 1 - 2 \gamma \right ) I 
+ \kappa_{es} \rho \int_{4 \pi} d \Omega' \sum _{n=0}^{3} \left ( { 2n + 1 
\over 4 \pi } \right ) P_{n} \left ( \xi \right ) S_n  I \cr 
& - {3 \kappa_{es}\rho \over 16 \pi} {c^2 \over h \nu^3 } \gamma I 
\left ( 1 - \nu \der { \, }{\nu} \right ) \int_{4 \pi} d \Omega' 
\left [ 1 - \xi + \xi^2 - \xi^3 \right ] I'  \, , \cr} \eqno (25) 
$$ 
where $P_{n}$ is the Legendre polynomial of order $n$ and 
$S_n$ ($n = 0,\dots,3$) are second order differential operators 
(see Pomraning 1973). 
Using the standard relation
$$
\xi = \mu \mu' + \sqrt { 1 - \mu'^2} \sqrt { 1 - \mu^2} 
\cos \left ( \Phi - \Phi' \right ) \, ,
$$
the previous expression can be cast into the form
$$
\eqalign{
{g_{C} \over E } = &  
\kappa_{es} \rho \left [ A_1 + \mu A_2 + \left ( 1 - \mu^2 \right ) A_3
+ \mu \left ( 3 - 5 \mu^2 \right ) A_4 \right ]
 - \kappa_{es} \rho f \left \{ 1 - 2 \gamma + \right.\cr
& \left.\tau 
\left [ A_5 - \mu^2 A_6 + \mu \left (3 \mu^2 - 5 \right )A_7 
+ \mu \left ( 3 - 5 \mu^2 \right ) A_8 \right ] \right \} \, .   
\cr } \eqno(26)
$$
The quantities $A_{i}$, containing the first four moments of $f'$ 
and their first and second frequency derivatives, are reported in Appendix A. 
This is the expression of $g_{C}$ needed in the general 
relativistic transfer equation in Fraser's approximation. We 
stress that up to now no assumptions have been made 
about the angular dependence in the energy exchange terms.      
A further simplification can be introduced if all terms, but $f$, in equation 
(26) are assumed to be isotropic and are replaced with their zero--th moments. 
The Compton source term becomes then
$$
\eqalign{
{g_{C} \over E} = & \kappa_{es} \rho j_{\nu} 
\left \{ 1 - \gamma + \gamma \der{\ln J_{\nu}} {\ln \nu}
+ \tau \left [ \derss{\ln J_{\nu}}{\ln \nu} + 
\left (\der{ \ln J_{\nu}}{\ln \nu} \right )^2 
\right. \right. \cr
 &  \left. \left. - 3
\der{\ln J_{\nu}}{\ln \nu} \right ] \right \} 
- \kappa_{es} \rho f \left [ 1 - 2 \gamma + { 1 \over m_e \nu^2} 
J_{\nu} \left ( 1 - \der{\ln J_{\nu}}{\ln \nu}
\right ) \right ] \, , \cr} \eqno(27) 
$$
where
$$
J_{\nu} = {1 \over 2} \int_{-1}^{1} I d \mu \, ,
$$
is the mean intensity.
The approximated expressions (26) and (27) 
are to be preferred whenever a non--relativistic plasma is considered, 
since their evaluation is much faster than that of the general 
source term given by equation (22). Moreover, equation (27) contains far fewer 
terms than (26), and has the great advantage that 
all the angular dependence is contained in $f$.

All forms of the Compton source term based on the Fokker--Planck 
approximation contain both first and second 
frequency--derivatives of the moments of the distribution function.
As noted by Nobili, Turolla \& Zampieri (1993), in connection with
the system of the first two PSTF moment equations, Compton terms act as 
singular perturbations, changing the mathematical character of the differential
operator that becomes elliptic. 
As we discuss in detail later on, our numerical code is based on an 
iterative scheme in which integral terms, together with their derivatives,
are treated as forcing terms, the only full--fledged differential operator 
being the one contained in the Boltzmann equation.
On the other hand, the characteristic ray method provides 
the angular and frequency dependence for $f$ that allows to write 
the Compton source term in its original form without resorting to
the Fokker--Planck approximation. In this case the problem of radiative 
transfer with comptonization can be solved exactly in any range of 
energies and optical depths, and 
the hyperbolic character of the Boltzmann equation is naturally preserved. 

\beginsection 4. THE NUMERICAL METHOD

In this section we describe in some detail the numerical scheme we have
developed for solving the transfer problem. The more general case, which
corresponds to spherical flows in a Schwarzschild spacetime, is discussed in 
subsection a); in subsection b) a simplified version of the code, for the 
solution of the full radiation hydrodynamical problem in static, 
plane--parallel atmospheres is presented; finally subsection c) is devoted to 
the numerical evaluation of the Compton source term.

\medskip
\centerline{\it a) The spherical case}
\medskip

As it is well known, in a scattering medium, the transfer 
equation is an integro--differential equation, while it has a simple
structure when only true emission--absorption is included; in particular, it
reduces to an ODE when written in its characteristic form.
This suggests that its solution can be found using an iterative method 
in which the starting point is just the solution of the transfer problem 
with only free--free processes taken into account. Following this 
idea, equation (12) has been integrated numerically, with the boundary 
conditions previously discussed, for a given set of values of the parameters 
$b^2$ and $E_{\infty}$ and with the source term $g = g_{ff}$.  
This provides the zero--th order approximation, $f^{(0)} (r, \mu, E)$ 
of the distribution function, that can be used 
to evaluate the scattering integrals appearing in $g_{es}$ or $g_{C}$. 
In the second step, we use the full expression for $g$ to obtain the first 
order approximation $f^{(1)} ( r, \mu , E )$. This is the solution of the
transfer equation written in the form
$${y(\mu + v)\over\rg}\dert{f^{(1)}}{r} = {{g_{ff}}\over E} + 
\alpha [f^{(0)}] - \beta [f^{(0)}] f^{(1)}\, .$$
All the expressions of the scattering source term discussed in the previous 
section can be cast, and have been presented, 
in this form. In equations (15), (16), (26) and (27) 
$\beta$ can be immediately identified with the 
coefficient of $f$; in equation (22) $\alpha$ is the integral term.      
The scheme is iterated until convergence is reached, 
improving at each iteration the functionals $\alpha$ and $\beta$
making use of the distribution function computed in the previous step.
As a convergence test, we compared each element of the matrix $j_{\nu}$ 
with its value relative to the previous iteration and stored the maximum 
relative correction. Cauchy criterion has been applied to verify the 
convergence of the succession of such corrections. 
 
Equation (12) has been integrated using a a finite differences method 
originally 
developed by Nobili \& Turolla (1988), in which the algebraic system is 
iteratively solved using the Henyey technique for matrix manipulation. 
The entire radial domain $[r_{in},r_{end}]$ is divided by $M$ points;
rays of type a) are integrated using this grid. For 
trajectories which exhibit a turning point, the transfer equation is solved
on the same mesh, picking up the subset of grid points which cover their
region of existence. Although, as we already mentioned, the choice of $r$
as the parameter along the geodesics has a number of advantages, it results in
a divergent derivative of $f$ at $\mu = -v$. While, for those branches which 
approach the turning point, this introduces some errors at most 
in the last few points, trajectories moving away from the
turning points may be systematically affected by an inaccurate determination of
their boundary condition. However, it should be taken into account that when 
the optical depth at the turning point is either large or very small, 
$f$ tends to $B_\nu/hcE^3$ or remains vanishingly small, independently on the
boundary condition for equation (12). Numerical errors, if any, are, then,
restricted to rays inverting in regions of moderate optical depth. 

The choice of the $b$--grid strongly constraints the final angular 
resolution of $f$, and requires special care. Let us first 
assume that a black hole is the central source; in this case 
the interval $0 \leq b^2 < 27/4$ corresponds to 
ingoing and outgoing trajectories of class a). 
For these two subclasses, we fix $N_1$ and $N_2$ 
values of the impact parameter in such a way to produce an equally--spaced
$\mu$--grid, in the range $[-1,1]$, at the critical point $r=3/2$.
To discretize the range $b^2\geq 27/4$, we exploit 
the one--to--one correspondence between $b^2$ and the position of the turning 
points, $r=r_n$, 
$$
b^2_n = {r_n^3 \over r_n -1 }\,.  \eqno(28)
$$
We fix $N_3$ and $N_4$ values of $r_n$, the first
at $r \leq 3/2$ and the latter
at $r \geq 3/2$; the $r_n$'s are just the radii of the 
spherical shells tangent to the orbits of types b) and c). 
In such a way, the total number $N$ 
of $\mu$--points in the interval $-1 \leq \mu \leq 1$ is $r$--dependent, 
and it is bounded by $N_1 + N_2 +1 \leq N \leq N_1 + N_2 + 2N_3 -1$ for 
$r \leq 3/2$ and $N_1 + N_2 + 1 \leq N \leq N_1 + N_2 + 2 N_4 -1 $ 
for $r \geq 3/2$. A better angular resolution in all the radial 
domain can be obtained increasing the number of photon trajectories.
In the case the central source is a star of radius $r_*$, the $\mu$--gridding
works in a very similar way, but the values of $b$ in the range 
$$
0 \leq \ b^2 < { r_*^3 \over r_* - 1 } 
$$     
now produce an equally--spaced $\mu$--mesh  at the star radius, the $N_4$ 
points refers to $r > r_*$ while no trajectory of type b) is present. 
We have found more convenient to derive the values of the impact parameter 
starting from the radial coordinate of the turning points, 
and not vice versa, since
in this way the radial extent of the photon trajectories, and hence the 
integration range of equation (12), is specified without solving the cubic 
equation (28). 

Once the rays are fixed, equation (12) must be integrated for different values
of the parameter $E_\infty$ along each trajectory. 
The range of $E_\infty$ should be chosen in such a way that, at each value
of $r$, we can compute the distribution function in an interval of the 
local energy, 
$[E_{min}, E_{max}]$, large enough to cover the interesting portion of the 
spectrum. The parameter range $[(E_\infty)_{min},(E_\infty)_{max}]$ must be
larger than $[E_{min}, E_{max}]$ at any given radius, since both gravity and 
dynamics act in changing the photon energy along the geodesics.
For $r<r_{end}$, in fact, 
the energy interval $[E_{min}, E_{max}]$ is actually 
influenced by some characteristic rays starting at $r_{end}$ with 
$E_{\infty}$ in the range
$$
(E_\infty )_{min} = \left [ y \left ( 1 + v \right ) \right ]_{r_{min}} 
E_{min} \leq E_{\infty} \leq 
\left [ y \left ( 1 - v \right ) \right ] _{r_{min}} E_{max} = 
(E_\infty)_{max} \, .
$$
In the numerical calculations we have used 
the dimensionless energy $x = E/KT_*$, where $T_*$ is a 
suitable normalization temperature. For later applications, we found more 
convenient to divide the storage window  
$[x_{min}, x_{max}]$ by means of $L$ points 
equally--spaced in $\ln x$;  
the same grid is maintained at all radii and $f$ is stored at these points 
as a function 
of the local dimensionless energy using an interpolation. 
In the two remaining ranges $[(x_\infty)_{min}, x_{min}]$ and 
$[x_{max}, (x_\infty)_{max}]$, $2P$ values of $x_\infty$ has been specified. 
For these values of the energy at infinity, the transfer equation has been 
integrated only along the trajectories of those photons that, at some $r$, 
have a local energy within our storage window.  
Loading the matrix $f(r_i,\mu_j,E_k)$ is particularly convenient since it
allows a more direct calculation of the scattering integrals, that are 
evaluated at both constant $r$ and $E$. All angular integrals can be obtained
simply performing a weighted sum of $f$ over the $\mu$--index without 
any additional scanning of the array or extra interpolations. 
This has, also, the advantage that the we are free to choose the most
suitable numerical scheme to integrate over energies since the rearrangement
of the energy points at each radius follows automatically.
The numerical evaluation of the frequency--dependent moments 
of $f$ has, however, to be carried out with some care. In particular, when the 
optical depth drops below unity and radial streaming is approached, the 
integration over $\mu$ becomes troublesome and we 
found more convenient to perform the quadrature over $b^2$, using equation
(11a). Since the same change of variable works well near the horizon, 
where outgoing rays concentrate towards $\mu =1$, it has been used 
in all the radial range. However, because of the divergence of 
$d \mu /d b^2$ where $\mu = - v$, in a small region around this point the 
original $\mu$--integration was performed at each value of $r$.

\medskip
\centerline{\it b) The Static, plane--parallel case}
\medskip

The numerical scheme we have just presented allows the solution 
of the transfer equation along the geodesic rays in the more general 
case, when gravity, dynamics and sphericity are all accounted for.
In many  astrophysical applications, however, transfer of radiation through
a static, geometrically--thin atmosphere is of interest, like, for example,
when studying reprocessing of thermal radiation in the atmosphere of 
X--ray bursting neutron stars. In all these cases, a plane--parallel approach
to the solution of the transfer problem is fully justified since the
atmospheric scale height is much less than the star radius, although the
effects of the strong gravitational field must still be considered. 
The assumption of hydrostatic equilibrium introduces a major simplification in
the treatment of radiative transfer because advection and aberration are no
more present. For a vanishing velocity field, equation (9) reduces to $E=
E_\infty/\sqrt{-g_{00}}$, implying that the value of the local energy at 
a given radius is the same along all rays. This is just another way of
stating the existence of Thorne's (1981) Universal Red--shift Function.
The rays are now symmetrical with respect to the $\mu=0$ line.
A further, drastic, simplification follows if it can be assumed that the
radial coordinate is constant in the atmosphere and equal to the star radius.
This is commonly done in non--relativistic transfer theory, replacing 
the height above the base of the atmosphere with the optical depth. 
The rays are just straight lines, $\mu = const$,
while the photon energy seen at infinity is simply the energy at any depth 
red--shifted by the constant factor $(1-1/r_*)^{1/2}$. In the light of these
considerations, we have developed and tested a reduced version of our code
which uses the scattering depth as the independent variable. The angular
mesh is obtained specifying directly the values of $\mu$; the energy
points at which $f$ is computed coincide with the energy grid, 
which is the same
at all depths. The calculation proceeds exactly in the same way as in a
non--relativistic problem and the spectrum at infinity is simply obtained 
by applying the gravitational red--shift factor to the spectrum emerging 
at the top of the atmosphere. 

An application to isolated neutron stars accreting at low rates is
presented in section 5b. In this problem electrons are far from being 
relativistic so Comptonization can be safely treated in the diffusion 
approximation using expression (27) for the scattering source term. The much 
shorter computational time allowed us to solve also the thermal and pressure 
structure of the atmosphere, coupling the hydrostatic balance and the radiative 
energy equilibrium to the transfer equation. 
The hydro equations are solved iteratively, exploiting the scheme for the
computation of the scattering integral we have already discussed.
Pressure and temperature profiles are computed at each iteration step, 
once the frequency--integrated moments have been obtained.

\medskip
\centerline{\it c) Numerical evaluation of the Compton source term}
\medskip

As discussed in section 3, the Compton source term, in the form (22), is the
sum of two contributions. The second addendum, which requires the 
calculation of the zero--th moment of the CSK, $\sigma_{00}$, poses no problems
since it involves a single quadrature of an analytical function. As proposed
by Shestakov {\it et al.\/} (1988), upon the change of variable
$$
u = { 1 \over \sqrt{2 \tau } } \left ( \sqrt{ z} - {1 \over \sqrt{z}}
\right ) \, ,$$
$\sigma_{00}$ can be efficiently evaluated using a Gauss--Hermite quadrature.
We have tested that six points give an accuracy better than 3 parts in 1000, 
sufficient for our purposes. 

We are left, then, with the problem of finding a fast algorithm for the 
numerical calculation of the multiple integral
$$
\int_0^{\infty} d \gamma' \left [  \exp \left (-{\gamma - 
\gamma' \over 2} \right ) \left ( {f \over 2} + 1 \right ) - {f \over 2} 
\right ] \int_{4\pi} d \Omega' 
\sigma \left ( \gamma \rightarrow \gamma', \xi , \tau \right ) f'\,. \eqno (29)
$$
First of all, we note that the scattering probability may become strongly 
peaked; in the Thomson limit, for example, the CSK tends toward a 
$\delta$--function at $\gamma = \gamma'$. In all regimes in which the integrand
is fastly--varying particular care must be used to account for delicate 
cancellations between opposite terms. We start considering the CSK itself. 
As discussed by Kershaw, Prasad and Beason (1986), the complicated 
three--dimensional integral in the electron velocity space can be reduced to 
a single integral when the solid angle element is defined with respect to a 
particular polar axis. In fact, taking the polar axis in the direction of 
the photon momentum transfer ${\vec s} = ( \gamma' {\vec n'} - 
\gamma {\vec n})/q $, $q = \sqrt{ \gamma^2 + \gamma'^2 - 
2 \gamma \gamma' \xi}$, and using the Dirac $\delta$--function to 
integrate over the polar angle, the integration over the azimuthal angle 
becomes analytical. The final form of the CSK is then
$$
\eqalign {
& \sigma  ( \gamma \rightarrow \gamma', \xi , \tau  ) =  
{3 \over 32 \gamma \nu \tau K_2 \left ( 1/ \tau \right )} \exp \left ( - 
\lambda_+ / \tau \right ) \left \{ { 2 \gamma \gamma' \tau \over q} \right . 
\cr & \left .
+ \int_{\lambda_+}^{\infty} \exp \left ( - { \lambda - \lambda_+ \over \tau} 
\right ) \left \{ { 1 \over \left ( 1 - \xi \right )^2 }   \right . \right . \cr
& \left . 
\left . \times \left [ { \left ( \lambda + \gamma \right ) \left ( { 1/
\gamma } + { 1/\gamma' } \right ) - \left ( 1 + \xi \right ) \over \left
[ \left ( \lambda + \gamma \right )^2 + \omega^2 \right ]^{3/2} } 
+ { \left ( \lambda - \gamma' \right ) \left ( { 1/
\gamma } + { 1/ \gamma' } \right ) + \left ( 1 + \xi \right ) \over \left
[ \left ( \lambda - \gamma' \right )^2 + \omega^2 \right ]^{3/2} } \right ] 
\right . \right . \cr
& \left . \left .
 + \left [ - \gamma \gamma' + { 2 \over 1 - \xi } + { 2 \over \gamma 
\gamma' \left ( 1 - \xi \right )^2 } \right ] \right. \right . \cr
& \left . \left . \times \left [ \left [
\left ( \lambda + \gamma \right )^2 + \omega^2 \right ]^{-1/2}  -
\left [ 
\left ( \lambda - \gamma' \right )^2 + \omega^2 \right ]^{-1/2}  \right ]
\right \} d \lambda \right \}  \cr} \eqno(30)
$$
where the Lorentz factor $\lambda$ is now the integration variable,
$\omega^2 = \left ( 1 + \xi \right ) / \left ( 1 - \xi \right )$ and 
$$
\lambda_+ = { \gamma ' - \gamma \over 2} + \left \{ \left [ 1 + \gamma \gamma'
{ 1 - \xi \over 2 } \right ] \left [ 1 + { \left ( \gamma - \gamma' \right )^2
\over 2 \gamma \gamma' \left ( 1 - \xi \right ) } \right ] \right \}^{1/2}\,.
$$

As stressed by Kershaw {\it et al.\/}, the main features of the scattering
probability are contained in the $\exp(-\lambda_+/\tau)/q$ term:
everything is smoothly varying with respect to this quantity, in particular 
with respect to the exponential. Kershaw {\it et al.\/} proposed two 
methods for the numerical evaluation of the $\lambda$--integral in equation
(30); in both cases the CSK is reduced to an approximate analytical 
expression. Here we adopt their fastest, although less accurate, algorithm
which is based on a suitable division of the integration domain into 
subintervals where the exponential is replaced by a linear interpolation.
To avoid delicate cancellations when $\tau\to 0$, a Taylor expansion
of the inner expression in curly brackets is used to obtain an asymptotic 
series in terms of Legendre polynomials for the integral; only terms up to 
second order are retained. Using this method the evaluation of 
the CSK becomes analytical with an accuracy of about 3 parts in a thousand
in all parameter ranges. The CPU time for a single 
evaluation is typically few microseconds on an alpha DEC--3000.  

The algorithm  we adopt for computing integrals involving the CSK 
follows the original method presented by Kershaw (1987) for evaluating the
total scattering cross--section and it is based on 
the fact that $\lambda_+$ has a minimum in both $\gamma'$ and $\xi$.
The most important contribution to the CSK comes, in fact, from regions near 
this minimum; everywhere else the scattering probability goes to zero 
exponentially fast with an $e$--folding length that is simply $\tau$ in
$\lambda_+$. Having these considerations in mind, 
the double angular integral in expression (29) can be written,
taking $\xi$ and $\overline\phi$ as the polar and azimuthal angles, as
$$
\int_{4 \pi} d \Omega' \sigma \left ( \gamma \rightarrow \gamma', \xi, \tau 
\right ) f' = \int_{-1}^1 d \xi \sigma \left ( \gamma \rightarrow \gamma' , 
\xi , \tau \right ) \int^{2 \pi}_0 d \overline \phi 
f \left ( r, \mu', \gamma' \right ) \, .    
$$
For each value of $r$, $\mu$, $\gamma'$, $\xi$ the azimuthal integral is 
evaluated using a Lobatto quadrature. The values of the distribution function 
at 
$$
\mu'_{l} = \mu \xi + \sqrt{ 1 - \mu^2} \sqrt{ 1 - \xi^2} \cos \overline 
\phi_{l}
$$
where $\overline \phi_{l}$ are the Lobatto abscissae, are obtained from a
linear interpolation. Once this is done, for each value of 
$\gamma$, $\gamma'$, $\tau$, the integration over all polar directions is 
carried out picking up the $\xi$ range, within the interval $\left | \xi 
\right | \le 1 $, that provides a non--negligible contribution to the 
scattering probability: as we anticipated, this is the region around the 
minimum of $\lambda_+$. For fixed $\gamma$, $\gamma'$ and $\tau$, the 
$e$--folding lengths in $\lambda_+$, $n\tau$, immediately provide the 
$e$--folding lengths in $\xi$, $\xi_n$. Denoting, in fact,
with $\xi_m=1 - |\gamma - \gamma'|/(\gamma\gamma')$ the value of $\xi$ 
where $\lambda_+$ is minimum, $\xi_n$ is the root of the equation 
$$
\lambda_{+m1} + n \tau = \lambda_+ \left ( \gamma, \gamma',
\xi_n \right ) \, ,
$$
where 
$\lambda_{+m1} = \lambda_+ \left ( \gamma, \gamma', \xi_{M} \right )$ and
$$
\xi_{M} = \max \left ( \min \left ( \xi_{m}, -1 \right ), 1 \right )
\, .
$$ 
Within each $e$--folding interval we use a 4--points Lobatto quadrature and the 
number of intervals is fixed by the request that either the
fractional contribution of the last $e$--folding is less than the desired 
accuracy ($5 \times 10^{-3}$ in the present case) or the boundary $\xi=\pm 1$ 
is reached. At $\gamma = \gamma'$ and $\xi = 1$ the CSK has an integrable 
($\sim\sqrt{1 - \xi}$) singularity, that can be easily eliminated
with the change of variable $\eta = \sqrt { 1 - \xi}$. 
  
The integration in energy is carried out in a similar way. Since 
the most important contribution to the inner integral (over $\xi$ in our 
scheme) comes from regions where $\lambda_+$ is near 
$\lambda_{+m1}$, the larger contribution to the outer integral
(over $\gamma'$) is provided by regions where $\lambda_{+m1}$ itself is 
minimum. Clearly, the lowest values of 
$\lambda_{+m1}$ correspond to $\xi_{M} = \xi_{m}$, i.e. to
$\xi_m \geq -1$; the inequality $\xi_m \leq 1 $ is always satisfied. 
We distinguish two cases: for $\gamma' \leq \gamma$ the previous 
condition is verified in the interval 
$$
{\gamma \over 1 + 2 \gamma} \leq \gamma' \leq \gamma
$$
that we call region A, while for $\gamma' \geq \gamma$ it holds 
in two different domains, that we call in general region B, depending on the 
value of $\gamma$:
$$
\eqalign{
\gamma \leq \gamma' \leq { \gamma \over 1 - 2 \gamma}~~~&{\rm if} ~~~\gamma < 
1/2 \cr
\gamma \leq \gamma' < \infty ~~~&{\rm if}~~~\gamma \geq 1/2 \, .\cr} 
$$
Since in region A $\lambda_{+m1} = 1 $, the search for the $e$--folding 
lengths is not required and integration is straightforward. 
This is not the case in region B, where $\lambda_{+m1} = 1
+ \gamma' - \gamma$. Now, although its minimum value is still $\lambda_{+m2}
=1$, $\lambda_{+m1}$ is {\it not\/} a constant. 
The corresponding $e$--folding lengths $\gamma'_n$ are to be derived solving 
the equation
$$
\lambda_{+m2} + n \tau = \lambda_+ \left ( \gamma, \gamma'_n ,\xi_M 
\right ) \, , \eqno(36)
$$     
which reduces to
$$ 
1 + n \tau = 1 + \gamma'_n - \gamma \eqno(37) 
$$
and gives simply $\gamma'_n = \gamma + n \tau$. 
In region B integration over $\gamma'$ is carried out using the same
procedure introduced for the $\xi$--quadrature.
To complete our discussion, we need to consider the two intervals 
$$
0 \leq \gamma' < {  \gamma \over 1 + 2 \gamma}\, ,
$$
region C, and, if $\gamma < 1 /2$, 
$$
{ \gamma \over 1 - 2 \gamma } \leq \gamma' < \infty \, ,
$$
region D. 
In both cases $\lambda_{+m1} = \lambda_+ \left ( \gamma, 
\gamma', -1 \right )$ and its minimum is reached at
$$
\lambda_{+m2} = \left . \lambda_{+m1} \right |_{\gamma' = \gamma / \left ( 
1 + 2 \gamma \right ) }
$$
or 
$$
\lambda_{+m2} = \left . \lambda_{+m1} \right |_{\gamma ' = \gamma 
/ \left ( 1 - 2 \gamma \right ) }
$$
in regions C and D respectively. 
The corresponding $e$--folding lengths are obtained from equation (36).
Lobatto rule is used everywhere and stepping is terminated when
its fractional contribution becomes less than $ 5 \times 10^{-3}$.
The most convenient number of Lobatto points depends on the typical 
relative values of the photon energy and gas temperature. In fact, for
different values of 
$\gamma$ and $\tau$, $\lambda_+(\gamma')$ can be either strongly 
peaked or very broad near its minimum. Optimization requires some numerical
experimenting, looking for the best agreement between the direct evaluation of
the CSK double integral and $\sigma_{00}$ computed using equations (24).
For the test model presented in subsection 5b, we used either a six or a 
ten points quadrature. Accordance between the values of $\sigma_{00}$ 
obtained using the two methods is better than 3--4\%, with the larger 
errors in the external region where the radiation temperature (mean photon 
energy) is very far away from the gas temperature.
On the other hand, where the Compton parameter $Y_C$ (see e.g. Rybicki \& 
Lightman 1979) is greater than unity, accuracy is better than 7 parts in a 
thousand. 
We finally note that the choice of a gaussian--type quadrature is motivated,
basically, by the fact that we need to perform integrals of the CSK times
$f$. The distribution function must be interpolated to obtain its values
at the integration points. Clearly, gaussian--type quadratures with a fixed 
number of abscissae are much faster, although less accurate, than 
step--adaptive schemes, as the Simpson rule originally used by Kershaw.
Computational feasibility is also the reason for which we decided to evaluate
the integral (29) using the values of $f$ relative to the previous iteration.
Clearly, it is possible to rewrite expression (29) as
$$\eqalign{&
{f \over 2}\int_0^{\infty} d \gamma' \left[\exp \left (-{\gamma - 
\gamma' \over 2} \right )-1\right] \int_{4\pi} d \Omega' 
\sigma \left ( \gamma \rightarrow \gamma', \xi , \tau \right ) f'\cr
&+\int_0^{\infty} d \gamma' \exp \left (-{\gamma - 
\gamma' \over 2} \right ) \int_{4\pi} d \Omega' 
\sigma \left ( \gamma \rightarrow \gamma', \xi , \tau \right ) f'\, \cr}
$$
since $f$ does not depend on $\gamma'$. Now $f$ is just the dependent variable 
of the transfer equation at any iterative step. The drawback is that the
computing times is about doubled, because of the two multiple integrals. The
CPU time for a single evaluation of expression (29) is typically $\sim 0.1$ s
and, in an production run, a $\sim 2\times 10^5$ evaluation are required,
implying a total time of about 6 hr.

\beginsection 5. APPLICATIONS

As we stressed several times, accretion flows onto compact objects, and
black holes in particular, provide an ideal arena for applications of
relativistic radiation transfer in differentially--moving media. The
accreting matter reaches, in fact, not only $r\sim 1$ with 
$v \sim 1$, but, often, temperatures high enough ($T\magcir 10^9$ K) to 
make a full treatment of Comptonization necessary. For this reason we
decided to present in this section the numerical solutions of the transfer
problem relative to different accretion regimes onto black holes and
neutron stars: ``cold'' and ``hot'' black hole accretion is considered in
subsections a) and b), respectively, and subsection c) deals with ``cold'',
static atmospheres around neutron stars. The full radiation hydrodynamical
problem in solved only in the latter case, while in the first two examples
the flow hydrodynamics is kept fixed. Since our present goal is to test the
capabilities of our method, no attempt has been made to explore the models
parameter space: we just present results for a single model which we judge 
useful in illustrating the main features of our integration scheme. 
For model c) a direct comparison with the results obtained by Zampieri
{\it et al.\/} (1995) with the moment expansion has been made, showing
a good agreement. No previous solutions for black hole accretion spectra 
are available, at least for models which contain an optically thick core.
Our attempt to cross-check results presented in section 5a integrating
the moment equations were hindered by severe numerical problems which
arise when the flow is not effectively thick at the horizon at all 
frequencies. The moment method can not, also, be used to compute radiative 
transfer in ``hot'' models,
where Compton scattering must be treated outside the Fokker--Planck 
approximation.
\medskip
\centerline{\it a) Accretion onto black holes: low--luminosity solutions}
\medskip
Spherical, stationary accretion onto a Schwarzschild black hole has been
throughly investigated in the past and we refer to the paper by Nobili, Turolla
and Zampieri (1991, NTZ in the following) for all details. A distinctive 
feature of black hole accretion is that, for the same value of the accretion 
rate $\dot m$ which is the only free parameter, two solutions may exist with 
very different properties: a ``cold'', low--luminosity and a ``hot'', 
high--luminosity one. In both solutions the sonic point is so far away that
we can safely assume that matter is free--falling with $u^1 = r^{-1/2}$
in our region of interest. Here we refer to models with high accretion rates, 
$\mdot\magcir 1$ in Eddington units. In this regime low--luminosity solutions 
start to develop an inner region optically thick to both 
free--free and scattering and show negligible Comptonization;
consequently, electron scattering can be treated in the Thomson limit, 
using expression (16). Under these conditions we expect, however, bulk motion
Comptonization in the converging flow (Blandford \& Payne 1981; Payne \&
Blandford 1981; Nobili, Turolla \& Zampieri 1993) to act efficiently
at high frequencies where true absorption is very low.

For our first test we consider NTZ solution characterized by $\mdot = 0.71$, 
corresponding to a density at the horizon $\rho_H = 10^{-6} \, {\rm g\, 
cm^{-3}}$. The gas temperature is in the range $2\times 10^4 \ {\rm K}\mincir 
T\mincir 5\times 10^5$ K, so we have chosen a normalization temperature
$\ln T_* = 11$. The dimensionless energy window is $x_{\min} = 0.1< x< 
x_{max} = 40$, corresponding to the range 0.5--206 eV; here $L=30$ points 
have been used. Outside this range, 
$P= 10$ energies has been fixed in each of the two additional 
intervals of $x_{\infty}$ we need, as discussed in section 3a. Since 
the effective optical depth at our larger energies is everywhere $<1$,
we solved the transfer problem for $10^{-2} \leq \log r \leq 5$, 
imposing the boundary condition $df/dr = 0$ along trajectories starting 
at $r_{in}$. The radial domain has been divided by $M = 250$ points; the
grid is not uniformly spaced and points are tighter around $r=3/2$.
To obtain a good angular resolution, 90 trajectories have been followed at 
each energy, $N_1 = N_2 = 20$, $N_3 = 10$, $N_4 = 40$. In such a way, the
number of $\mu$ points, which is minimum at $r=3/2$, is always greater
than 21. At each value of $x$ within the storage window 
the scattering source term was calculated using a linear interpolation 
for both the matrices $j_{\nu}$ and $k_{\nu}$; outside this window an 
extrapolation has been used. 
Figures 2a and 2c show the mean intensity $J_{\nu}$ and radiation pressure 
$K_{\nu}$ at different energies, together with the Planck function at 
$T(r_{in})$; each curve corresponds to a different value of 
the radial coordinate. The effective frequency--dependent 
optical depth goes from $3 \times 10^{3}$ to $10^{-4}$ for the 
lowest frequency, while high energy photons stream freely at all radii. 
The low energy portion of the spectral distribution is a superposition of 
thermal bremsstrahlung emission at different temperatures while
bulk motion Comptonization produces a power law high--energy tail. The 
calculated spectral index, $\alpha = -2.9$, is due to unsaturated 
bulk motion comptonization, being the scattering optical depth $\sim 0.7$ at 
the horizon. The 
theoretical value derived by Payne \& Blandford, in the limit $\tau_{es} 
\gg 1 $, is $\alpha = -2$ for a free--fall
velocity. At large radii, where radial streaming is approached, all moments 
fall off as $r^{-2}$; the asymptotic radial gradient we have found is
$-1.99$. While the 
evaluation of even moments does not pose particular 
problems, in the inner regions, where the radiation field is nearly 
isotropic, a direct numerical quadrature for computing odd moments becomes 
troublesome because of the delicate cancellations 
between contributions of opposite sign. To avoid this problem, the 
monochromatic flux, presented in figure 2b, has been replaced with its 
analytical expression in the diffusion approximation every time it is 
$\tau_{eff} > 10$. A typical production run required 
10--11 iterations to converge with a fractional accuracy better than 
$10^{-4}$, with a total CPU time of about 20 minutes on an alpha 
DEC--3000.

\medskip
\centerline{\it b) Accretion onto black holes: high--luminosity solutions}
\medskip

In ``hot'' solutions temperature is much higher, typically $\sim 10^{10}$ K 
near to the horizon. As a consequence, free--free absorption is much lower
than in ``cold'' models, even for larger accretion rates. Along the 
high--luminosity branch, thermal Comptonization is the dominant radiative 
process and it must be treated in its more general form, using expression (22).
Here we consider the ``hot'' solution of NTZ with  $\mdot = 71$, $\rho_H = 
10^{-4} \, {\rm g\, cm^{-3}}$. The flow is now effectively thin at all 
frequencies, although an inner core optically thick to scattering is present. 
The gas temperature in this model is in the range 
$10^5 \ {\rm K}\leq T \leq 10^{10} \ {\rm K}$, so we have chosen $\ln T_* = 21$
and $x_{\min} = 0.008$. Now the energy window is 0.9--4500 keV and $L = 35$ 
points have been used. Since the evaluation of the CSK integrals is very
time--consuming, both angular and radial resolution has been reduced
with respect to the previous model: $N_1 = N_2 = N_3 = 10$, $N_4 = 30$ and 
$M= 110$ in the same radial domain. In this model convergence has been 
reached with a fractional accuracy better than 0.02, and the calculated 
radial gradient at infinity is $-2.04$.    
The resulting mean intensity is presented in figure 3. 
In high temperature models, the mean intensity is always less than $B_\nu$, but
despite the accreting gas radiates less efficiently than in low--luminosity
optically thick solutions, the efficiency of accretion process is higher, due 
to the fact that the matter temperature is now higher 
in the whole photospheric 
region. Since the emergent spectrum is peaked 
at about 40 keV, 
these solutions, if stable (see NTZ and Zampieri, Miller \& Turolla 1995), 
seem to provide a natural way to produce hard 
X--ray radiation with reasonable efficiency out of spherical accretion onto 
black holes. 
     
\medskip
\centerline{\it c) Static, plane--parallel atmospheres around 
neutron stars}
\medskip

Our last application refers to a static, ``cold'' atmosphere around an 
accreting unmagnetized neutron star. Emitted spectra were firstly 
derived by Zel'dovich \& Shakura (1969) and, in more detail, by Alme \& Wilson 
(1973) in the high luminosity range ($l \magcir 10^{-3}$ in Eddington units). 
Solutions
for $10^{-7} < l < 10^{-3}$ has been recently presented by Zampieri 
{\it et al.\/} (1995) in connection with isolated neutron stars 
accreting the interstellar medium. As they have shown, the emitted spectrum 
exhibits an overall hardening with respect to the blackbody at the neutron 
star effective temperature with a hardening ratio, typically $\sim 1.5-3$,
which increases with decreasing luminosity.
The most important physical processes are free--free emission and 
absorption; Compton cooling plays a role only in increasing  
the temperature in the outer atmospheric layers, 
where the low energy tail of the 
emitted spectrum is created. Since typical temperatures are very low, 
Comptonization is treated by means of the approximated expression (27),
which is much faster. The run of thermodynamical variables is
obtained solving the hydrostatic and energy balance (see Zampieri {\it et 
al.\/}) together with 
the transfer equation, using an iterative scheme. However, in this 
problem the thermal balance is very delicate and the zone where 
photons of different energies thermalize strongly depends on integrated 
quantities, $J$ and the absorption mean $\kappa_0$. Numerical integration 
proved more stable if $J$ and $H$ are derived as solution 
of the first two gray moment equations. 
The same approach was used by Zampieri {\it et al.\/}, 
with the difference that in our scheme 
the gray moment equations are solved exactly,
computing from the specific intensity the Eddington factors $K/J$ at each 
depth and $J/H$ at $\tau = 0$. 
Here we have recomputed the model with $l = 10^{-4}$, using a spectral 
window $x_{min} = 0.1$, $x_{max}=10$ centered around a normalization 
temperature $\log T_* = 6.6$; $L = 30$ frequency points have been used.  
The angular resolution is provided by $N_1=N_2=15$ trajectories and 
the transfer equation is solved in the range $ -8 <\log(\tau )< 0.9$ using 
$M = 100$ grid points. The resulting mean intensity is plotted in figure 4.
Figures 5 and 6 show the emergent spectrum and the temperature profile together
with the results of Zampieri {\it et al.\/} (dashed 
lines). As can be seen, the agreement both in the spectral shape and the 
temperature profile is very good, showing that the approximated solution of 
the transfer equation with two moments is rather accurate.
To obtain this model, with a fractional accuracy better than 
$2 \times 10^{-2}$, 14 iterations were required, with a 
total CPU time of about 3 minutes on an alpha 
DEC--3000. Agreement between the gray mean intensity, derived as the double 
integral of $f$, and the solution of the second gray moment equation is 
always better than few parts in thousand.  

\beginsection 6. CONCLUSIONS

In this paper we have presented a characteristic method for the solution 
of the general relativistic transfer equation. If the spacetime admits
some symmetries, the formalism can be simplified; in particular, 
in presence of three Killing vectors, two of the three equations 
for the characteristic rays become analytical. In addition, using the radial 
coordinate as the parameter along the null geodesics, the exact solution 
of the transfer problem can be obtained solving a single ordinary 
differential equation along a bi--parametric family of characteristic 
trajectories. A numerical technique, based on an iterative scheme, 
has been developed and tested either for the calculation of the radiation
field in a fixed background or for the solution 
of the full radiation hydrodynamical problem
in spherical and plane--parallel geometry.
Particular care has been devoted to the evaluation of the source term, taking 
into account radiative processes which are believed to be of importance in 
astrophysical accreting plasmas: electron--electron and electron--proton 
bremsstrahlung, Thomson and Compton scattering. 

Radiative effects due to magnetic fields and 
pair production--annihilation were not considered in this work. 
However we stress that 
the method we have presented 
is completely general and additional radiative processes 
may be easily included. 
Source terms not involving integrals 
of the photon distribution function can be simply accounted for 
when the corresponding emissivity and opacity coefficients are provided. 
On the other 
hand, our iterative scheme allows for the solution of integro--differential 
equations and can be used 
to include also different integral source terms as, for instance, those ones 
related to pair production or bound--bound emission. 
Actually, a self--consistent treatment of pair production entails  
the solution of the full radiation hydrodynamical 
problem, with the addition of the pair balance equation and was left out on 
purpose in our discussion of ``hot'' accretion solutions which are 
obtained at fixed hydrodynamics.  

In the test models we discussed 
magnetic fields and pair production are not expected to play 
a relevant role at least for low luminosity solutions. In the case of accretion 
onto neutron stars it can be easily shown, in fact, that, for typical 
temperatures and densities in the photospheric region, 
the cyclotron emission is lower than the free-free 
emission if $B \mincir 10^9$ G (see e.g. Schmid--Burgk, 1978). On the 
other hand, a relic magnetic field of this order 
is just what is expected in isolated neutron stars which evolved 
beyond the pulsar phase; our models can be then assumed to describe 
correctly the emitted spectrum from old neutron stars accreting the 
interstellar medium. As far as 
low--luminosity accretion onto black holes is concerned, 
the limiting value is a factor 
$10^{-2}$ smaller, but it still exceeds the maximum strength of the tangled 
$B$--field derived assuming equipartition between magnetic and thermal 
energy densities. 
However, in the inner regions  
of high--luminosity models electrons become 
relativistic and 
both pair processes and synchrotron emission start to play 
a role.  

\beginsection REFERENCES

\ref{Alexanian, M. 1968, Phys. Rev., 165, 253}
\ref{Alme, M.L., \& Wilson, J.R. 1973, ApJ, 186, 1015}
\ref{Anderson, J.L. \& Spiegel, E.A. 1972, ApJ, 171, 127}
\ref{Anile, A.M., \& Sammartino, M. 1989, Ann. Phys., 14, 325}
\ref{Anile, A.M., \& Romano, V. 1992, ApJ, 386, 325}
\ref{Blandford, R.D., \& Payne, D.G. 1981a, MNRAS, 194, 1033}
\ref{Blandford, R.D., \& Payne, D.G. 1981b, MNRAS, 194, 1041}
\ref{Castor, J.I. 1972, ApJ, 178, 779}
\ref{Chandrasekhar, S. 1960, Radiative Transfer, (New York: Dover)}
\ref{Cooper, G. 1971, Phys. Rev. D, 3, 2312}
\ref{Dermer, C.D. 1984, ApJ, 280, 328}
\ref{Dermer, C.D. 1986, ApJ, 307, 47}
\ref{Gould, R.J. 1980, ApJ, 238, 1026}
\ref{Haug, E. 1975, Zs. Naturforschung, 30a, 1099}
\ref{Hauschildt, P.H. \& Wehrse, R. 1991, J. Quantit. Spectros. Radiat. 
Transfer, 46, 81}
\ref{Heitler, W. 1936, The Quantum Theory of Radiation, 
(Oxford: Oxford University Press)}
\ref{Kershaw, D.S., Prasad, M.K., \& Beason, J.D. 1986, J. Quantit. Spectros. 
Radiat. Transfer, 36, 273}
\ref{Kershaw, D.S. 1987, J. Quantit. Spectros. Radiat. Transfer, 38, 347}
\ref{Kompaneets, A.S. 1956, Soviet Phys. JETP, 4, 730}
\ref{Levermore, C.D., \& Pomraning, G.C. 1981, ApJ, 248, 321}
\ref{Lindquist, R.W. 1966, Ann. Phys. (NY), 37, 487}
\ref{Loeb, A., McKee, C.F., \& Lahav, O. 1991, ApJ, 374, 44}
\ref{Madej, J. 1989,  ApJ, 339, 386}
\ref{Melia, F., \& Zylstra, G.J. 1991, ApJ, 374, 732}
\ref{Mihalas, D., Kunasz, P.B., \& Hummer, D.G. 1975, ApJ, 202, 465}
\ref{Mihalas, D., Kunasz, P.B., \& Hummer, D.G. 1976a, ApJ, 203, 647}
\ref{Mihalas, D., Kunasz, P.B., \& Hummer, D.G. 1976b, ApJ, 206, 515}
\ref{Mihalas, D. 1980, ApJ, 237, 574}
\ref{Mihalas, D., Winkler, K--H., \& Norman, M.L. 1984, J. Quantit. Spectros.
Radiat. Transfer, 31, 479}
\ref{Mihalas, D. \& Mihalas, B. 1984, Foundations of Radiation Hydrodynamics
(Oxford: Oxford University Press)}
\ref{Misner, C.W., Thorne, K.S., \& Wheeler, J.A. 1973, Gravitation, Box
25.7 (San Francisco: Freeman \& Co.)}
\ref{Nobili, L., \&  Turolla, R. 1988, ApJ, 333, 248}
\ref{Nobili, L., Turolla, R., \& Zampieri, L. 1991, ApJ, 383, 250}
\ref{Nobili, L., Turolla, R., \& Zampieri, L. 1993, ApJ, 404, 686}
\ref{Payne, D.G., \& Blandford, R.D. 1981, MNRAS, 196, 781}
\ref{Pomraning, G.C. 1973, The Equations of Radiation Hydrodynamics (New 
York: Pergamon Press)}
\ref{Pomraning, G.C. 1983, ApJ, 266, 841}
\ref{Prasad, M.K., Shestakov, A.I., Kershaw, D.S., \& Zimmerman, G.B. 1988,
J. Quantit. Spectros. Radiat. Transfer, 40, 29}
\ref{Quigg, C. 1968, ApJ, 151, 1187}
\ref{Rybicki, G.B., \& Lightman, A.P. 1979, Radiative Processes in 
Astrophysics (New York: Wiley)}
\ref{Schinder, P.J. 1988, Phys. Rev. D, 38, 1673}
\ref{Schinder, P.J. \& Bludman, S.A. 1989, ApJ, 346, 350}
\ref{Schmid--Burgk, J. 1978, APSS, 56, 191}
\ref{Shestakov, A.I., Kershaw, D.S., \& Prasad, M.K. 1988,
J. Quantit. Spectros. Radiat. Transfer, 40, 577}
\ref{Simon, R. 1963, J. Quantit. Spectros. Radiat. Transfer, 3, 1}
\ref{Stepney, S., \&  Guilbert, P.W. 1983, MNRAS, 204, 1269}
\ref{Thomas, L.H. 1930, Q. Jl. Math., 1, 239}
\ref{Thorne, K.S. 1981, MNRAS, 194, 439}
\ref{Turolla, R., \&  Nobili, L. 1988, MNRAS, 235, 1273}
\ref{Zampieri, L., Miller, J.C., \& Turolla, R. 1995, MNRAS, in the press}
\ref{Zampieri, L., Turolla, R. \& Treves, A. 1993, ApJ, 419, 311}
\ref{Zampieri, L., Turolla, R., Zane, S. \& Treves, A. 1995, ApJ, 439, 849}
\ref{Zel'dovich, Ya., \& Shakura, N. 1969, Soviet Astron.--AJ, 13, 175}

\beginsection APPENDIX A

Here we present the expressions of the $A_i$, $i = 1, \dots, 8$, 
terms appearing into equation (26). Their derivation starts from 
Fraser's result, equation (25), and makes use of relation
$$
\xi = \mu \mu' + \sqrt { 1 - \mu'^2} \sqrt { 1 - \mu^2} 
\cos \left ( \Phi - \Phi' \right ) \, .  
$$
At first order in $\gamma$ and $\tau$, only the first 
four moments $m_{\nu}^n$ of the distribution function 
$$
m_{\nu}^n = {1 \over 2} \int_{-1}^{1} f \mu ^n d \mu
$$ 
appear in the Compton source term; they 
will be termed $j_{\nu}$, $h_{\nu}$, $k_{\nu}$ and $l_{\nu}$, using the
standard notation.
We introduce also the correspondent moments of the specific intensity $I$: 
$$
M_{\nu}^n = E^3 m_{\nu}^n \, .
$$ 
To make the larger number of terms dimensionless, we introduce 
double logarithmic frequency derivatives for all even moments; terms
containing odd moments are written as semi--logarithmic derivatives since
odd moments may become vanishingly small in regions where the effective 
optical depth is very large.
The final result is:
$$
\eqalign{
A_1 = & j_{\nu} \left \{ 1 - \gamma \left ( 1 - \der{\ln \Jb }
{\ln \nu} \right ) + \tau  \left [ \derss{\ln \Jb}{\ln \nu} 
+ \left ( \der {\ln \Jb } {\ln \nu} \right )^2 - 3 
\der {\ln \Jb } {\ln \nu} \right ] \right \} \cr
A_2 = & {6 \over 5} \left [ \left ( \gamma - \tau \right ) h_{\nu} 
- { 1 \over E^3} \left ( \gamma - 3 \tau \right ) \der {\Hb}{\ln \nu}
- { \tau \over E^3} \derss{\Hb}{\ln \nu} \right ] \cr
A_3 =  & { 1 \over 8 } \left \{ \left ( 1 - \gamma - 6 \tau \right ) 
\left ( j_{\nu} - 3 k_{\nu} \right ) + \left ( \gamma - 3 \tau 
\right ) \left ( j_{\nu} \der{\ln \Jb} {\ln \nu} - 3 
k_{\nu} \der{ \ln \Kb} {\ln \nu} \right ) \right . \cr  
& \left. + \tau  \left [ j_{\nu} \left ( \left ( 
\der{ \ln \Jb}{\ln \nu} \right )^2 + \derss { \ln \Jb }{\ln \nu}
\right ) - 3 k_{\nu} 
\left ( \left ( \der{\ln \Kb}{\ln \nu} \right )^2 
+ \derss{\ln \Kb} {\ln \nu} \right ) \right ] \right \}
\cr
A_4 =  & { 3 \over 40} \left [ \left ( \gamma + 4 \tau \right ) 
\left ( 3 h_{\nu} - 5 l_{\nu} \right ) - { 1 \over E^3 } \left (
\gamma - 3 \tau \right ) \left ( 3 \der{\Hb}{\ln \nu} - 5 
\der { \Lb } {\ln \nu} \right ) \right . \cr
& \left . - { \tau \over E^3 } \left ( 3 \derss { \Hb } {\ln \nu}
- 5 \derss{\Lb} {\ln \nu} \right ) \right ] \cr
A_5 = & { 3 \over 8 } \left ( 3 j_{\nu} - k_{\nu} 
- 3 j_{\nu} \der{\ln \Jb } {\ln \nu} + k_{\nu}
\der{\ln \Kb} {\ln \nu} \right ) \cr
A_6 = & { 3 \over 8} \left ( j_{\nu} - 3 k_{\nu} - j_{\nu} \der{\ln \Jb} {\ln 
\nu} + 3 k_{\nu} \der{ \ln \Kb}{\ln
 \nu} \right ) \cr
A_7 = & { 3 \over 8 } \left ( h_{\nu} - { 1 \over E^3 } \der { \Hb } 
{\ln \nu} \right ) \cr
A_8 = & { 3 \over 8} \left ( l_{\nu} - { 1 \over E^3 } 
\der {\Lb} {\ln \nu} \right ) \cr}
$$

\beginsection FIGURE CAPTIONS

\refig{Figure 1a.\quad The run of the cosine of the angle between 
the photon momentum and the radial direction, as measured by 
a free--falling observer, along the characteristic rays. 
Different curves correspond to 
different values of the impact parameter $b$.}
\medskip

\refig{Figure 1b.\quad Same as in figure 1a for the 
photon energy normalized with respect to $E_\infty$.}
\medskip

\refig{Figure 2a.\quad 
Monochromatic mean intensity at different radii (full lines), 
together with the blackbody 
function at $T(r_{in})$ (dashed line), for ``cold'' 
accretion onto a black hole with $\mdot = 0.71$.}
\medskip

\refig{Figure 2b.\quad Same as in figure 2a for the monochromatic flux.}
\medskip

\refig{Figure 2c.\quad Same as in figure 2a for the monochromatic 
radiation pressure.}
\medskip

\refig{Figure 3.\quad 
Monochromatic mean intensity at different radii (full lines),  
together with the corresponding blackbody 
function at $T(r_{in})$ (dashed line), for ``hot'' 
accretion onto a black hole with $\mdot = 71$.}
\medskip

\refig{Figure 4.\quad 
Monochromatic mean intensity at different scattering depths 
(full lines) for a ``cold'', static, plane--parallel atmosphere around 
a neutron star with $l = 10^{-4}$. The blackbody 
function at $T(\tau_{in})$ is also drawn for 
comparison (dashed line).}
\medskip

\refig{Figure 5.\quad The emergent spectrum for the model in figure 4 
(full line) compared 
with the blackbody at the neutron star effective temperature 
(dash--dotted line) and with the solution obtained by Zampieri {\it et al.\/} 
(1995, dashed line).}
\medskip

\refig{Figure 6.\quad The gas temperature profile for the model in figure 4  
(full line), compared with that one found by Zampieri 
{\it et al.\/} (1995, dashed line).}

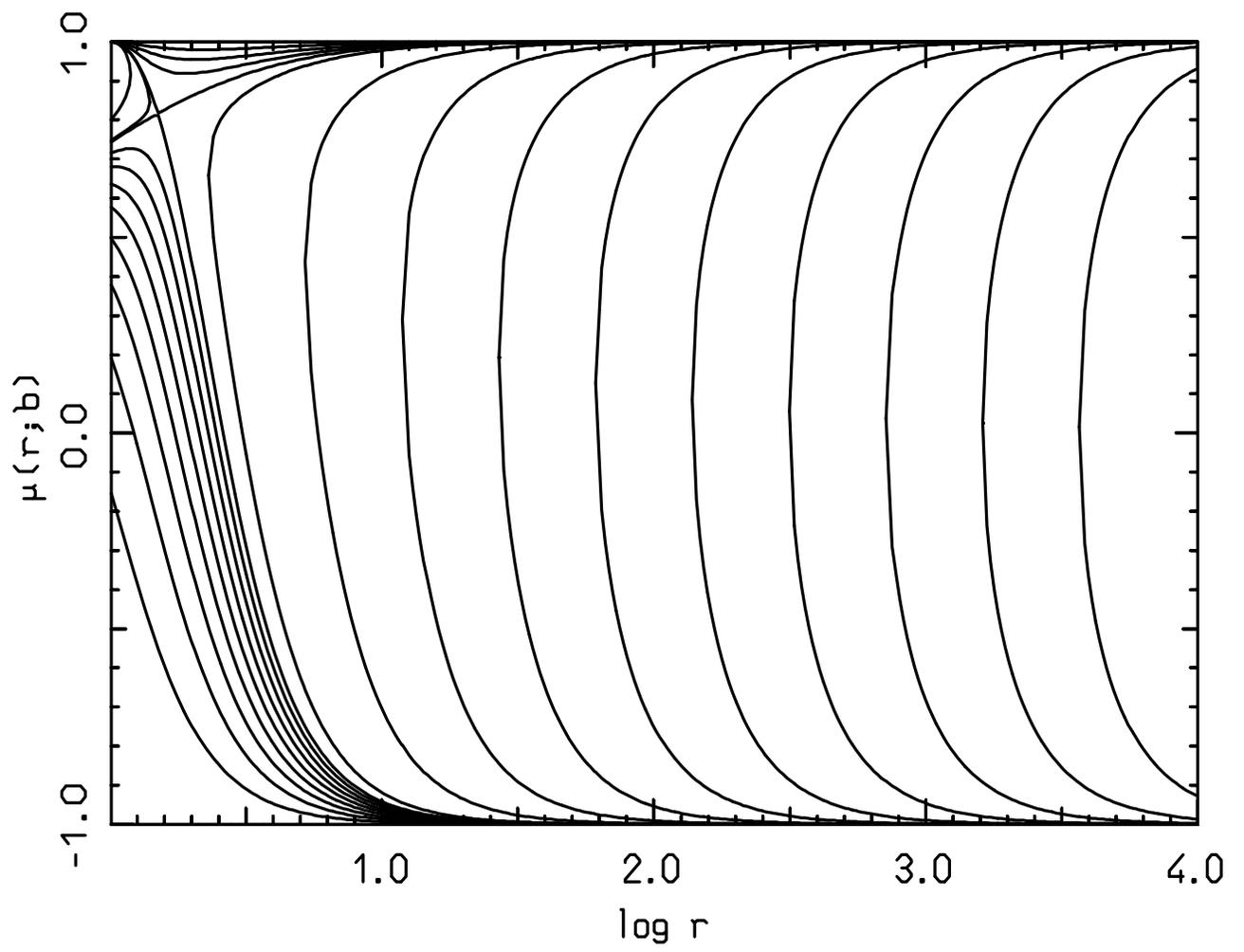
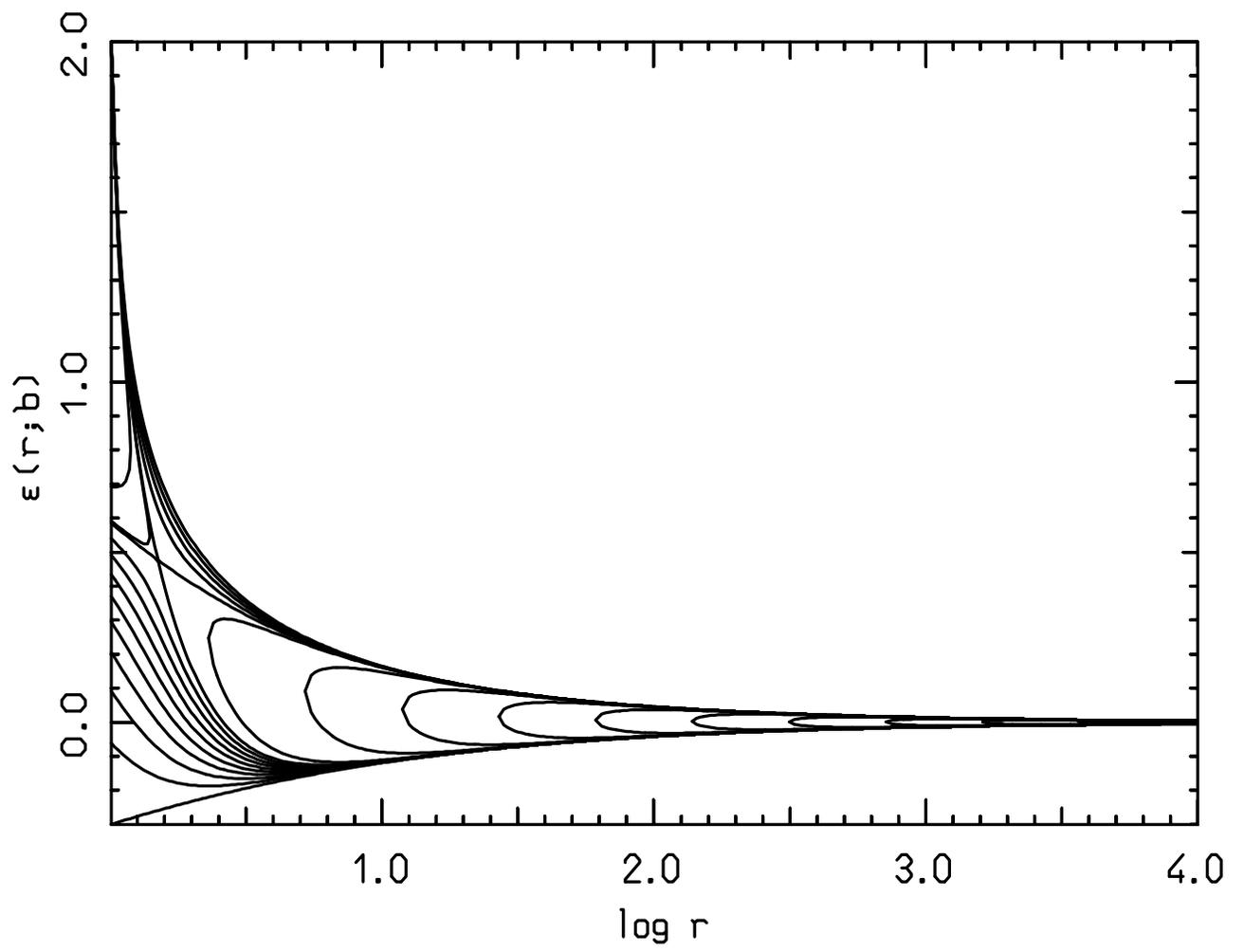
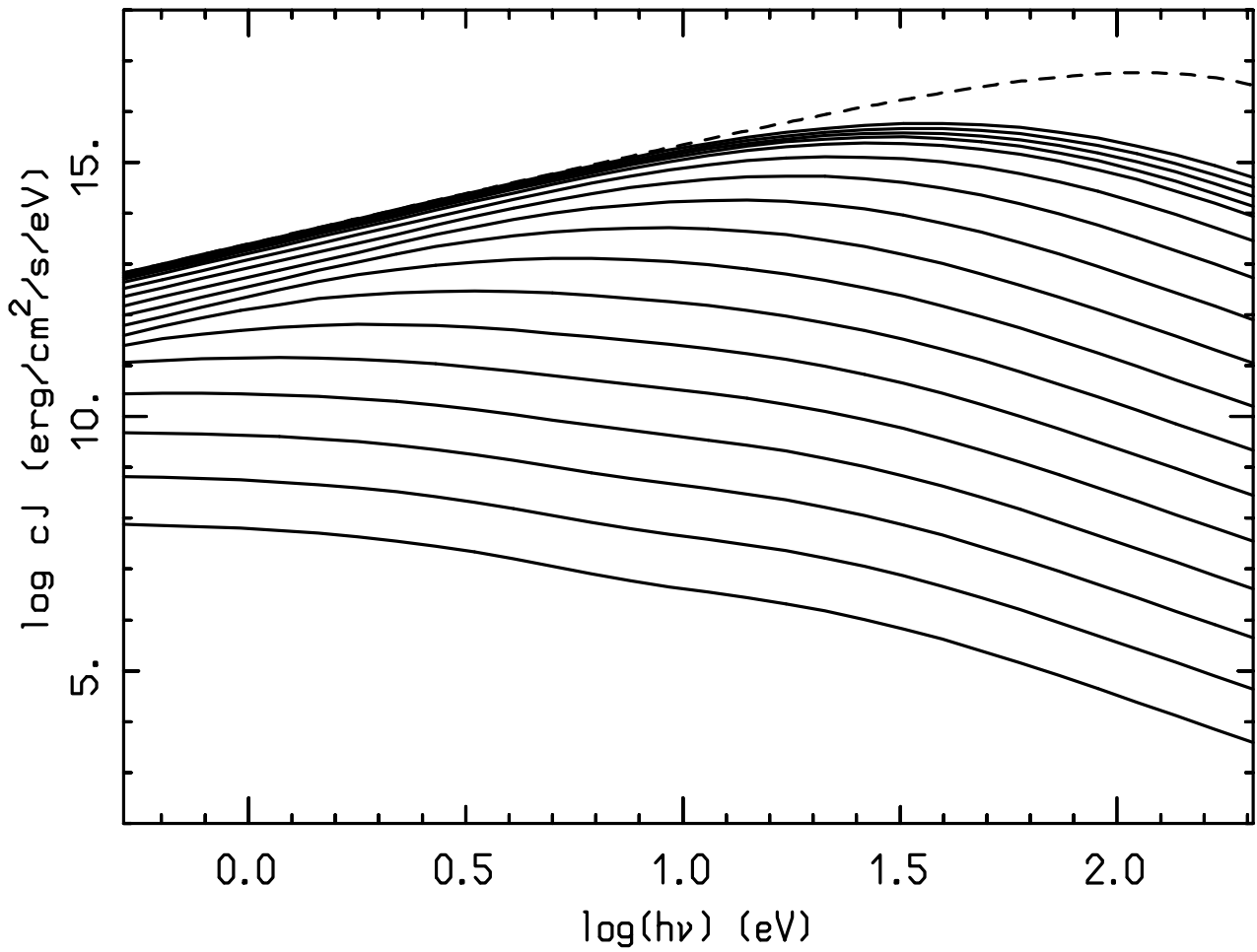
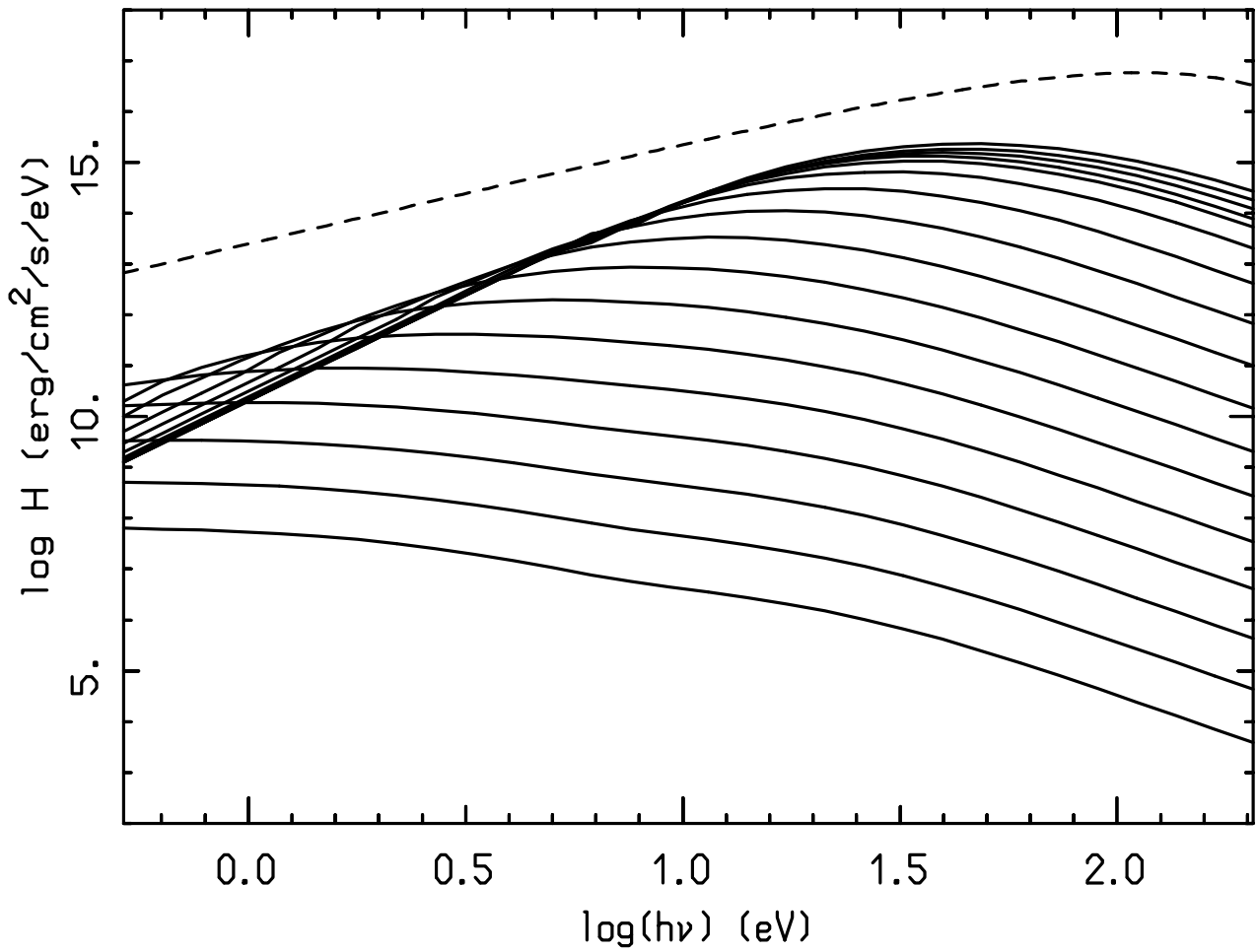
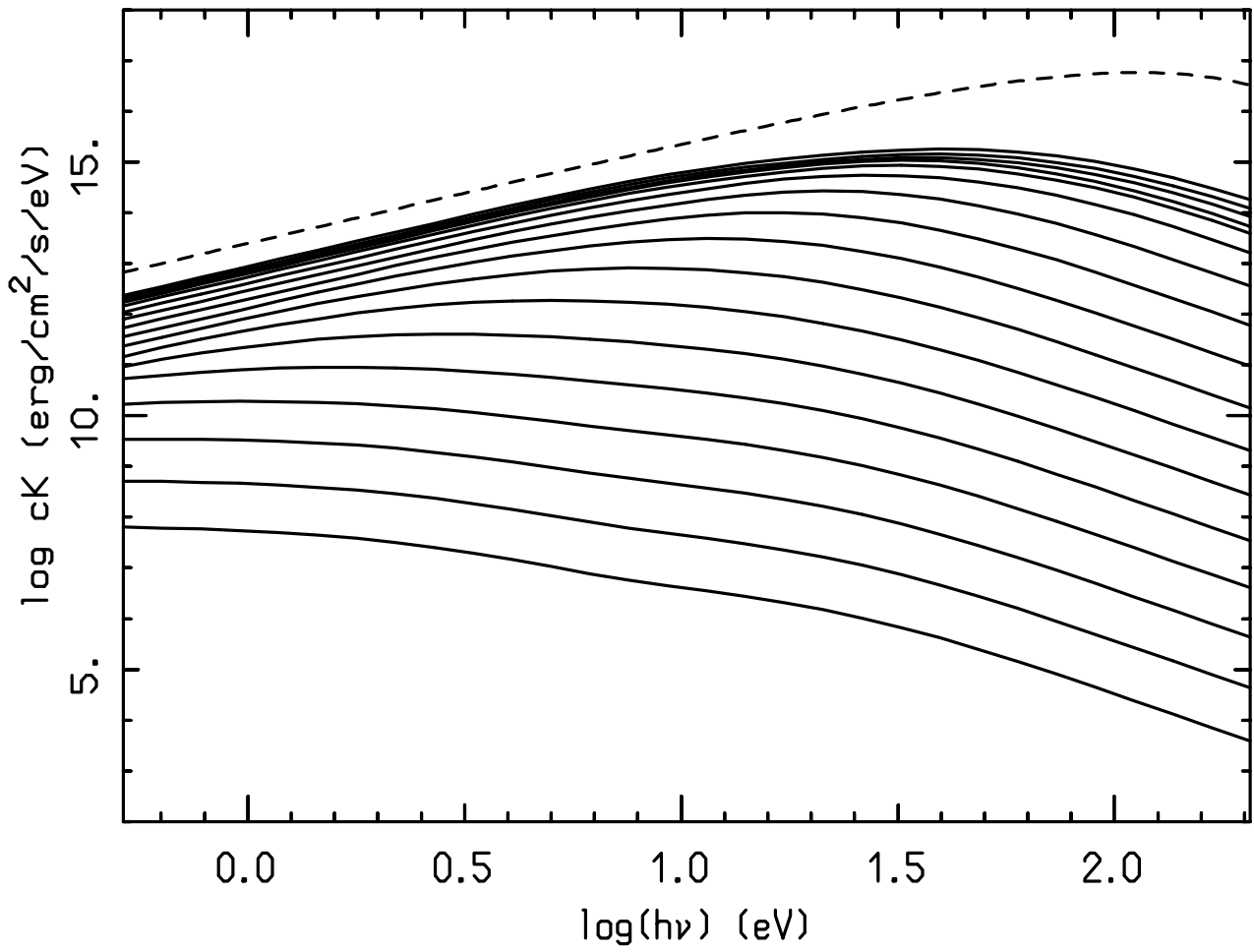
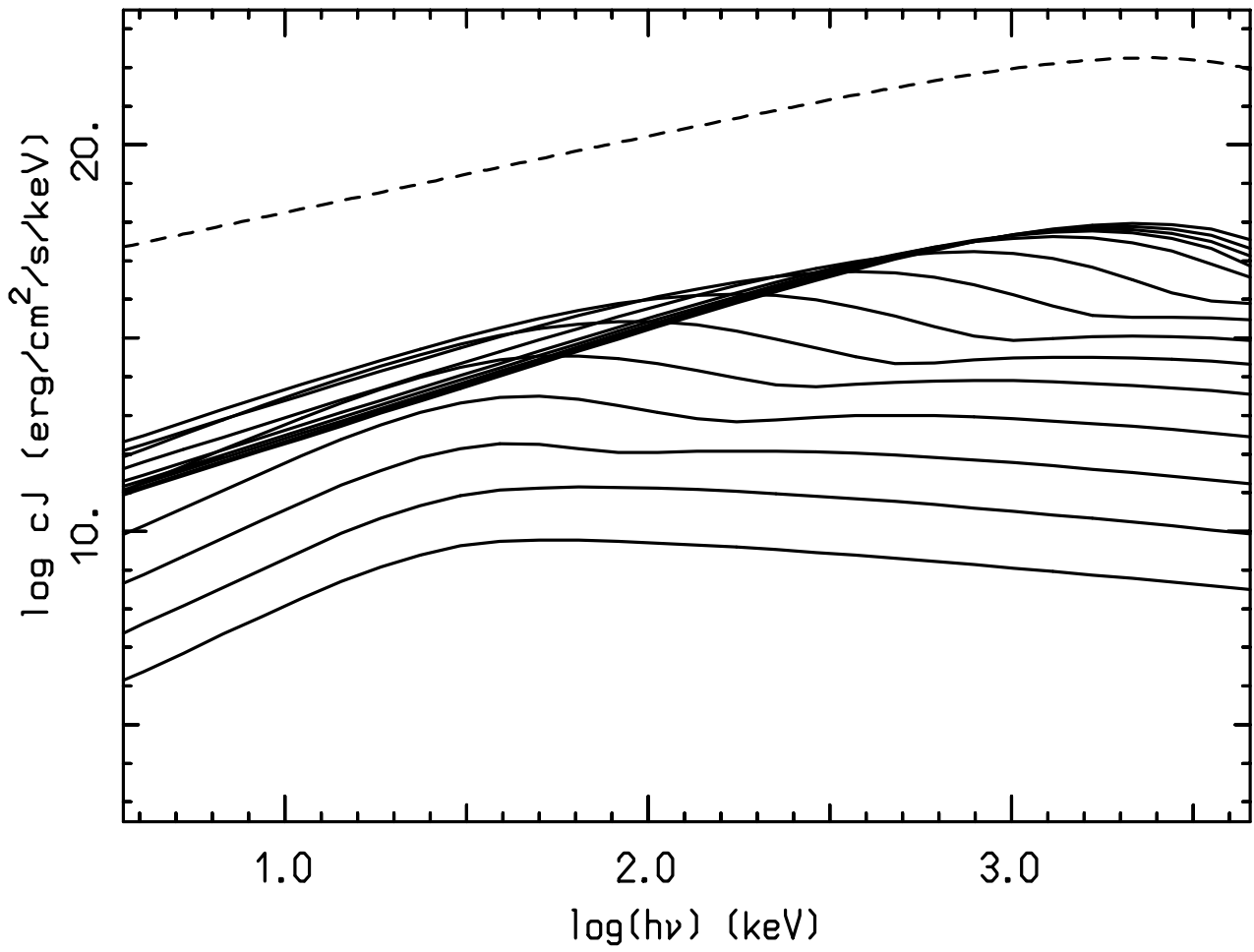
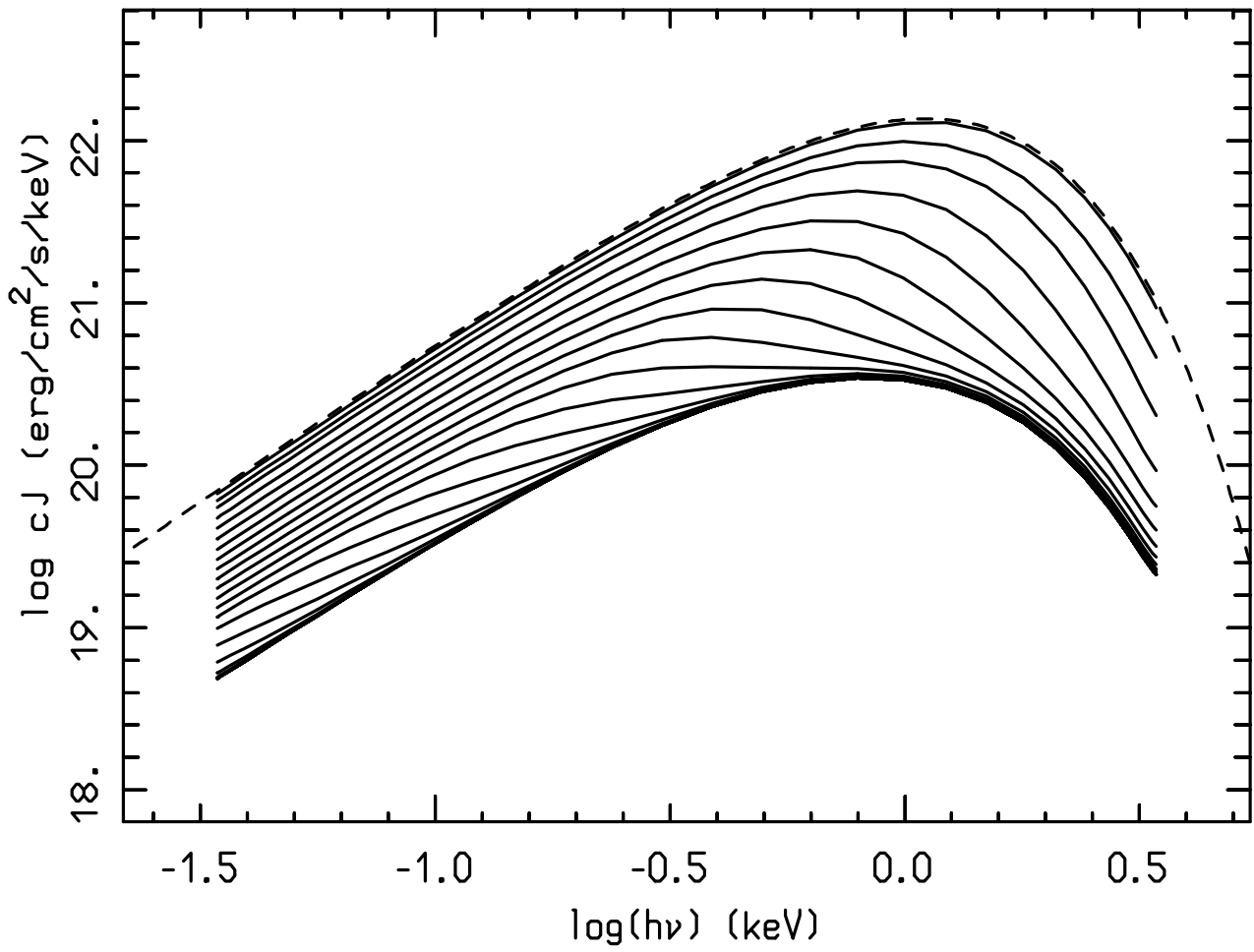
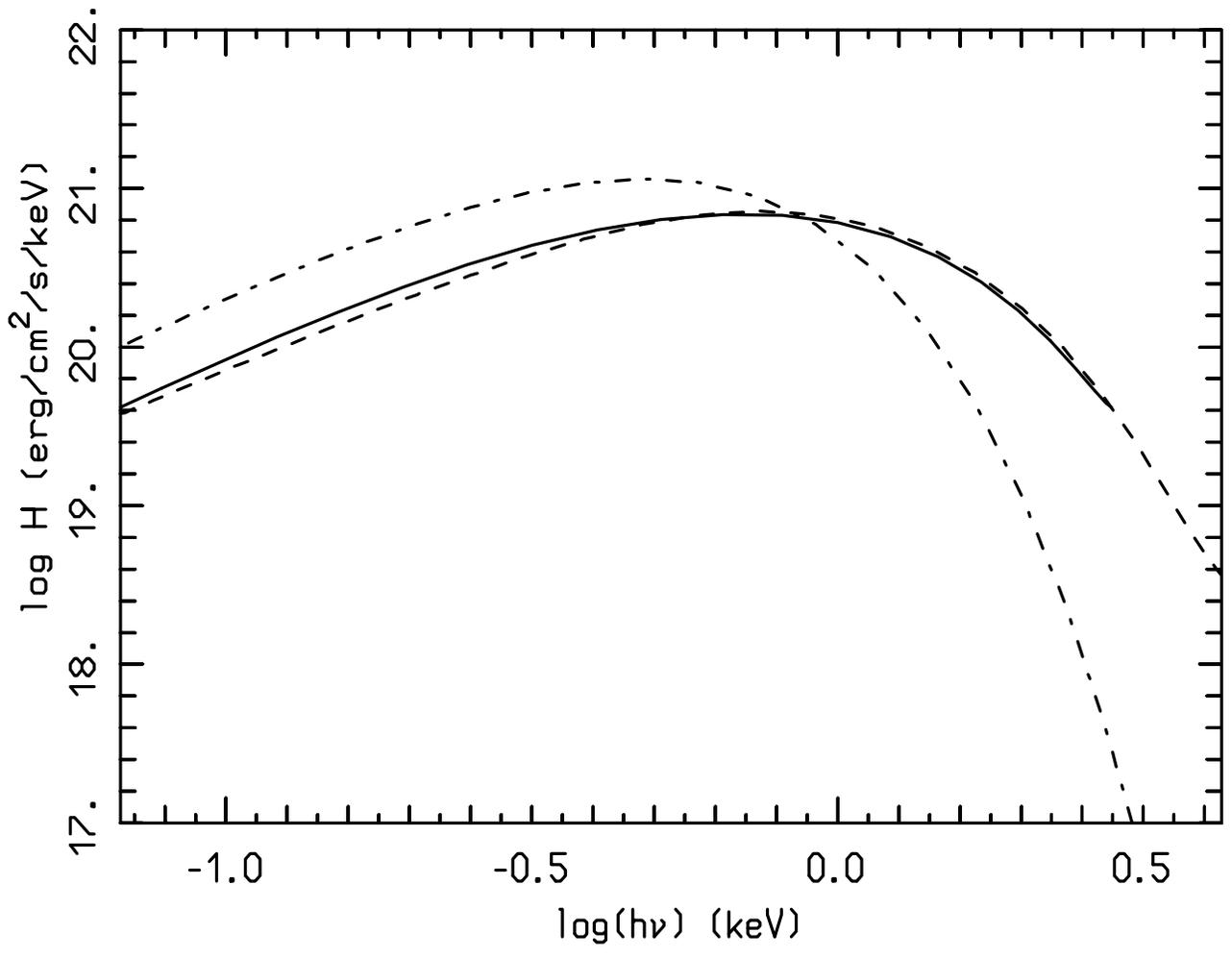
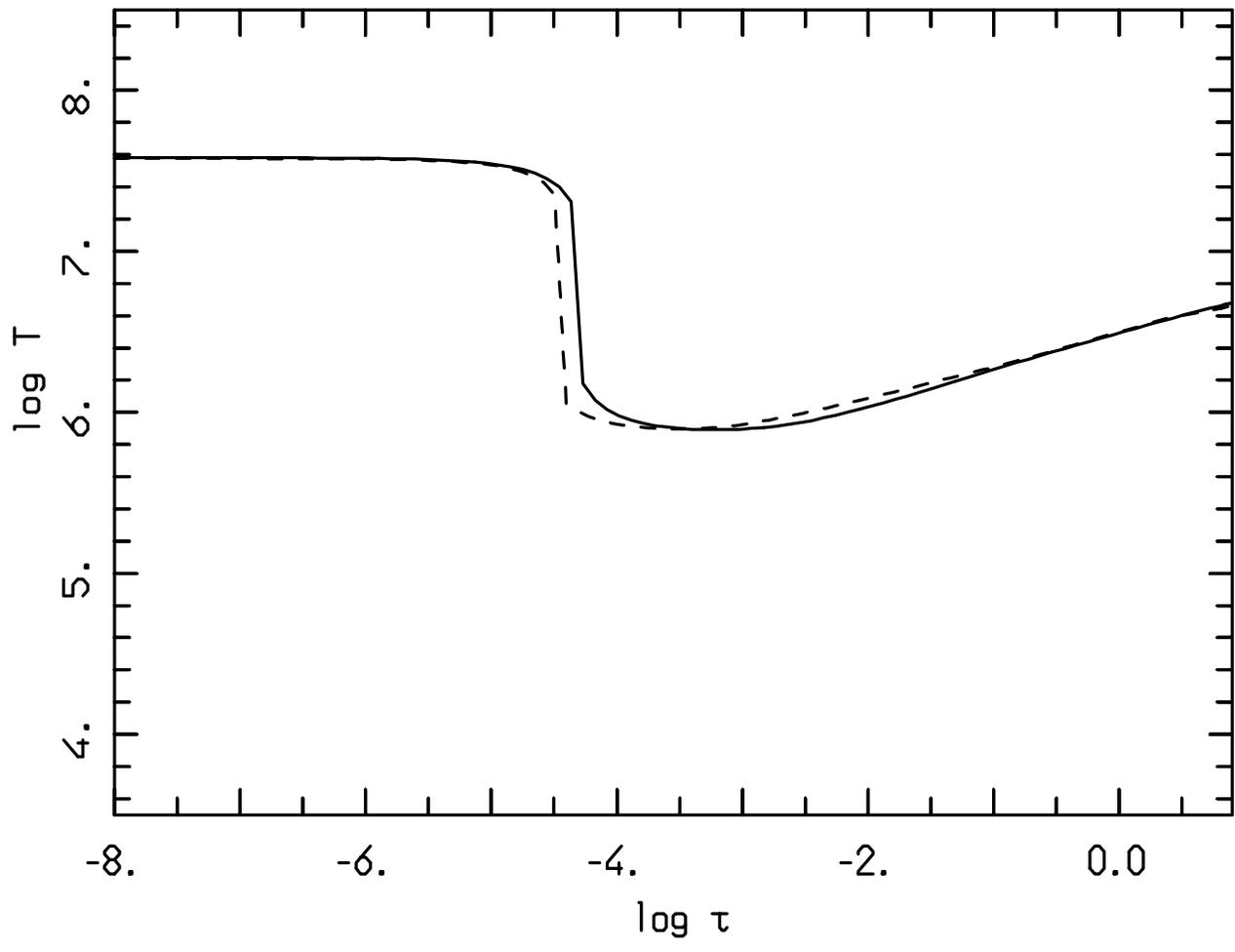
\bye